\documentclass[twocolumn,amssymb,prd]{revtex4}
\usepackage{graphicx}
\usepackage{dcolumn}
\usepackage{bbold}
\usepackage{amsmath}
\usepackage{amsfonts}
\usepackage{journal_names}
\usepackage{bm}
\usepackage{natbib}

\def\3nab{\tilde{\nabla}}

\def\hsp5{\hspace{5mm}}

\def\case#1/#2{\textstyle\frac{#1}{#2}}

\def\be {\begin{equation}}
\def\ee {\end{equation}}
\def\ber {\begin{eqnarray}}
\def\eer {\end{eqnarray}}
\def\bea {\begin{eqnarray}}
\def\eea {\end{eqnarray}}

\def\bc {\begin{center}}
\def\ec {\end{center}}
\def\case#1/#2{\frac{#1}{#2}}

\newcommand{\CD}{{\cal D}}

\renewcommand{\geq}{\geqslant}

\newcommand{\saverage}[1]{\left\langle #1 \right\rangle}
\newcommand{\average}[1]{\left\langle #1 \right\rangle_{\CD}}

\def\beq{\begin{equation}}
\def\eeq{\end{equation}}
\def\bea{\begin{eqnarray}}
\def\eea{\end{eqnarray}}
\def\ba{\begin{align}}
\def\ea{\end{align}}
\def\d{{\rm d}}

\begin{document}
\title[Averaging and backreaction in cosmology]{
Does the growth of structure affect our dynamical models of the universe?\\
\it -- The averaging, backreaction, and fitting problems in cosmology --
}
\author{Chris Clarkson$^1$, George Ellis$^1$, Julien Larena$^{2,1}$ and Obinna Umeh$^1$}
\address{$^1$ Astrophysics, Cosmology \& Gravity Centre, and, Department of Mathematics
and Applied Mathematics, University of Cape Town, Rondebosch 7701, South Africa
\\
$^2$ Department of Mathematics,
Rhodes University,
6140 Grahamstown,
South Africa
}

\date{\today}

\begin{abstract}
Structure occurs over a vast range of scales in the universe. Our
large-scale cosmological models are coarse-grained representations
of what exists, which have much less structure than there really is.
An important problem for cosmology is determining the influence the
small-scale structure in the universe has on its large-scale
dynamics and observations. Is there a significant, general
relativistic, backreaction effect from averaging over structure?
One issue is whether the process of smoothing over structure can
contribute to an acceleration term and so 
alter the apparent value of the cosmological constant.
If this is not the case, are there other aspects of concordance
cosmology that are affected by backreaction effects? Despite much
progress, this `averaging problem' is still unanswered, but  it cannot be ignored in an era of precision
cosmology, for instance it may affect aspects of Baryon Acoustic Oscillation observations.
\end{abstract}

\maketitle \setcounter{footnote}{0}
\DeclareGraphicsRule{.wmf}{bmp}{jpg}{}{}
\maketitle

\maketitle
\newpage
\section{Introduction}

A crucial feature of the real universe is that structure occurs on
many scales. Large-scale models of the universe, such as the
standard models of cosmology, are coarse-grained representations of
what is actually there, which has much more structure than is
represented by those models. An important issue for cosmology is the
question of whether the smaller scale structures influence the
dynamics of the universe on larger scales: is there a significant
back-reaction effect from the small scales to the large scales?

The standard Friedmann-Lema\^{\i}tre-Robertson-Walker (FLRW)
models of cosmology are a resounding success. They can account for
all observations to date with just a handful of parameters. These
are the simplest reasonably realistic universe models possible
within General Relativity: homogeneous, isotropic, and flat to a
first approximation, with a scale-invariant spectrum of Gaussian
perturbations from inflation added on top to account for structures down to the
scale of clusters of galaxies. Only the physical motivation for the
value of the cosmological constant and the associated coincidence
problem is required for a complete understanding at late times. Cosmology in
the future will be refining this picture.

But is it as simple as this? The universe may well be statistically
homogeneous and isotropic above a certain scale, but on smaller
scales it is highly inhomogeneous, quite unlike a FLRW universe.
General Relativity is a theory in which spacetime itself is the
dynamical field with no external reference space. Yet it is
ubiquitous in cosmology to talk of a `background' which is exactly
homogeneous and isotropic, on which galaxies and structure exist as
perturbations. Is this the same as starting with a more detailed
truly inhomogeneous  metric of spacetime, and progressively
smoothing it~-- probably by a non-covariant 
process~-- until we get to this background?

It is commonly assumed that whichever way one goes about it the
results of these two processes must be the same; but the
non-linearity of the field equations ensures that they are not. The
problem for cosmologists is whether this difference is important.
This has been the subject of controversy, with opinions ranging from
the suggestion it could completely explain the recently observed
accelerating expansion of the Universe without a need for any dark
energy
\cite{Buchert:1999er,Buchert:2001sa,Rasanen:2003fy,Barausse:2005nf,Kolb:2005da,Rasanen:2006kp,Kasai:2007fn}
to the claim that it is completely negligible
\cite{Ishibashi:2005sj,Flanagan:2005dk,Hirata:2005ei,Geshnizjani:2005ce,Bonvin:2005ps,Bonvin:2006en,
Behrend:2007mf,Kumar:2008uk,Krasinski:2009qq,
Tomita:2009ar,Baumann:2010tm,Green:2010qy}, with a middle-ground of
papers claiming it is enough to disturb the cosmic concordance of
the standard model
\cite{Russ:1996km,Kolb:2004am,Li:2007ci,Li:2007ny,Li:2008yj,Clarkson:2009hr,
Clarkson:2009jq,Chung:2010xx,Umeh:2010pr}.

Averaging has kinematic, dynamical, and observational aspects. One
estimate of the importance of this effect would be to note that
within the standard model, variations in the expansion rate of
around a few percent occur on scales of order
100\,Mpc~\cite{Russ:1996km,Wang:1997tp,
Shi:1997aa,Li:2007ci,Li:2007ny,
Clarkson:2009hr,Umeh:2010pr,Bolejko:2011ej}. At the very least,
then, the averaging problem may be important for observations in
precision cosmology; indeed it is obviously built into standard
observational practice.

But does it have dynamic consequences? Green and Wald
\cite{Green:2010qy} claim to derive in a rigorous and systematic way
the effects of small scale inhomogeneities on large scale dynamics,
and thereby prove that matter inhomogeneities produce no new effects
on the dynamics of the background metric: they can only mimic
radiation. However others reach an opposite conclusion: Buchert's
approach \cite{Buchert:1999er,Buchert:2001sa} shows backreaction
from inhomogeneity can potentially mimic dark energy;
 Kolb and colleagues concur \cite{Kolb:2011}, and even have suggested inhomogeneity effects can fully explain the
acceleration of the universe \cite{Kolb:2005me,Kolb:2005da}, 
as has Wiltshire~\cite{Wiltshire:2009ip,Wiltshire:2007jk}. Recently
Baumann et al.~\cite{Baumann:2010tm}, using second-order
perturbation theory, have claimed that virialised structures freeze
out of the cosmic expansion and only affect the background dynamics
through a renormalised mass.  Clearly there is no agreement, but even
from a conservative viewpoint there is evidence that backreaction
might have an impact on precision cosmology:  Baumann et al \cite{Baumann:2010tm}
conclude that \emph{inter alia} it significantly affects the Baryon
Acoustic Oscillations (BAO).

\subsection*{This review}
Our aim in this paper is to present the different approaches to the
dynamical aspects of backreaction effects in recent eras in cosmology,  
highlighting open questions and so future research agendas. While we
will briefly touch on them, we will not deal comprehensively with
observational aspects of the fitting problem: this raises a number
of further issues we do not have space to deal with
here \footnote{The effect of inhomogeneities on observations is
dealt with in detail in a forthcoming paper by Clarkson, Ellis,
Faltenbacher, Maartens, and Uzan.}. We will also not deal with
possible backreaction effects in the very early universe, which
raise a rather different set of issues (see e.g.
\cite{AbrBraMuk97,CalHuMaz01}).

 There have been quite a number of attempts to estimate the backreaction effect.
One set is based on starting with an inhomogeneous model and using
some averaging process to produce an approximately FLRW model. This
has been developed both in terms of general formalisms, and in terms
of detailed model building. Alternatively, by averaging over
structure in the standard model, the backreaction effect can be
calculated perturbatively; this should then provide an estimate of
its size, and, in principle,  allows us to  evaluate the fluctuations which become of the order of the mean, and the background one starts with no longer correctly describes the averages.

We first discuss non-perturbative attempts to estimate backreaction.
Then we provide an overview of backreaction in the standard model,
where it appears the effect is small~-- at least to second-order in
perturbation theory~-- but nevertheless is at least sufficient to affect
precision cosmology results.

\section{Averaging, Backreaction, and Fitting}

\subsection{The Lumpy universe and the Background}

\begin{figure*}[htbp]
\begin{tabular}{c}
\includegraphics[width=0.95\textwidth]{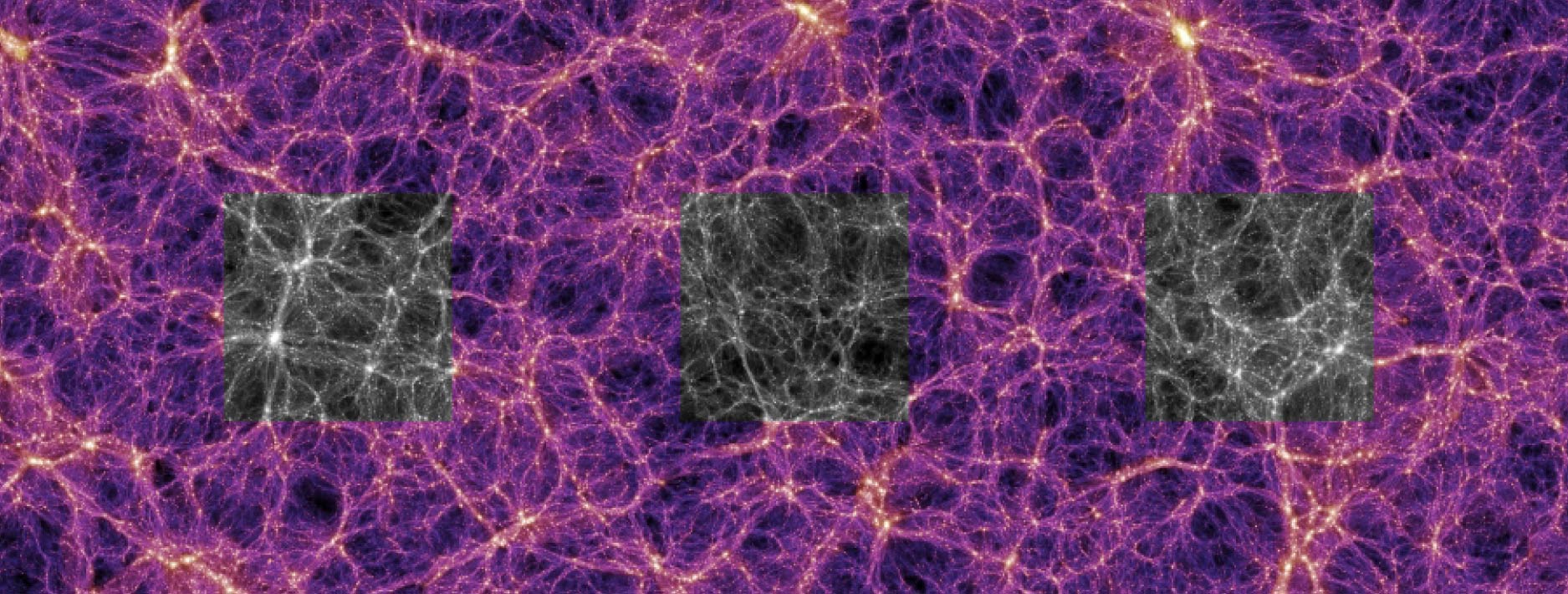}\\
\includegraphics[width=0.96\textwidth]{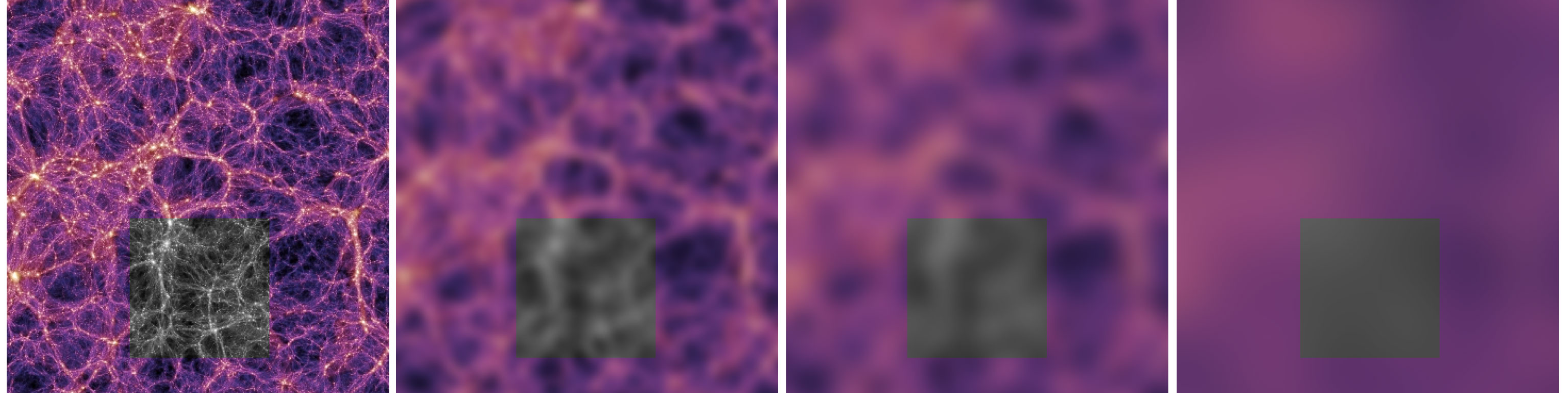}
\end{tabular}
\caption{Structure in the Millennium simulation~\cite{MilSim2005}
(from~\cite{Clarkson:2009jq}). Can we describe the universe as
smooth on scales of order $150$Mpc, shown here in the black and
white boxes (top panel)? The averaging problem is shown in the
bottom row: how do we go from left to right? Does this process give
us corrections to the `background', or is it the `background'
itself? How does it relate to the `background' left at the end of
inflation?} \label{fig:sim}
\end{figure*}

Any description of physical reality embodies a scale of averaging
that is usually not made explicit. The world looks completely
different depending on the scale of description chosen: for example
a fluid looks wholly different on an everyday scale (say 1 m) as
opposed to atomic scale (say $10^{-11}$m). The same is true for
astronomy and cosmology. The background models in standard cosmology
are the FLRW models, which ignore all local structural details. More
realistic models include linearised perturbations about this
background, in principle representing the largest scale growing
structures down to the scale of clusters of galaxies (10's of Mpc)
from early times to today; but they do not represent the non-linear
smaller scale structures, such as our galaxy or the solar system, at
later times. The universe looks far from homogeneous when viewed on
any scale from 1~AU to 1~Mpc, and only  approaches statistical
homogeneity past 100~Mpc (though even this is in
dispute~\cite{Labini:2010aj,Labini:2010mg})~-- see
Fig.~\ref{fig:sim}.

To describe the Universe on its largest scales, one has to make
approximations that postulate or derive a high degree of symmetries
for the metric of space-time. Practically, this means that one wants
to calculate the large scale observables using a background
geometry, i.e. a geometry that ignores, on average, the details
present on small scales and that are not probed by the observables.
Such a background is usually found in a FLRW solution to the
Einstein Field Equations (EFE). This issue is referred to as the
\emph{fitting problem}~\cite{EllSto87}: what is the best-fit FLRW
model to the lumpy Universe? In the standard concordance cosmology,
the existence of this background is postulated, and no smoothing
mechanism  is provided to obtain it from the real lumpy Universe.
Despite the success of this approach in fitting the observations
there still remains the problem of properly defining the background
geometry in relation to the real lumpy universe.

For example, which set of observers (worldlines) are associated to
the background, i.e. are supposed to measure an homogeneous and
isotropic Universe? It is clear that such a fitting procedure needs
an explicit method of averaging or smoothing. This averaging can be
performed with different techniques and on different quantities. For
example, one can argue that homogeneity is a spatial property and
average quantities that define particular spatial hypersurfaces,
such as densities, pressures, or expansion rate of geodesics
bundles. In that case, the background is defined by surfaces of
constant density, pressure or expansion rate and the average
quantities are used to fit the background. Another possibility is to
fit the model via averaged observable relations such as the
magnitude-redshift or number count-redshift relations; then the
average has to be performed in some sense on the past-null cone of the observer.
Of course, these procedures will, in general, give different results
and the FLRW fitting model reconstructed will depend on which method
has been used.

The construction of the background is a crucial issue: if the wrong
background is compared to the data, it will imply the existence of a
backreaction that may disappear if a better background is chosen.
This emphasizes the fact that a gauge choice is always part of an
averaging procedure. As such, all the approaches listed and
commented on in what follows somehow intend to clarify the (usually
unstated) way this is handled in the standard approach to
cosmological modeling in most papers on cosmology.

We have three closely related but distinct problems to consider:
\begin{description}
\item[Averaging] Coarse-graining of structure, such that small-scale
effects are hidden to reveal large scale geometry and dynamics.
\item[Backreaction] Gravity gravitates, so local gravitational inhomogeneities may affect
the cosmological dynamics. How this is calculated depends on the
degree of coarse graining.
\item[Fitting] How do we appropriately fit an idealized model to observations made from one location in a
lumpy universe, given that this `background' does not in fact exist?
\end{description}

\subsection{Averaging and Backreaction} The basic issue is the non
commutativity of averaging and the field equations. Start with a
realistic description of the universe on a small scale, with metric
$g^{\text{(local)}}_{ab}$ (e.g. this might describe individual stars
and planets in the universe, and the vacuum between them). Average
it by a smoothing procedure to a metric $g^{\text{(gal)}}_{ab}$ with
an averaging scale where galaxies are well represented but
individual stars are invisible. Average this in turn to a metric
$g^{\text{(lss)}}_{ab}$ with an averaging scale where large scale
structures are well represented but individual galaxies are
invisible (there are many possible scales that are omitted in this a
description). The largest scale (completely smoothed) model will
have a FLRW metric $g^{\text{(cos)}}_{ab}$, where all traces of
inhomogeneity have been removed.  There will similarly be averaged
stress energy tensors $T^{\text{(local)}}_{ab}$,
$T^{\text{(gal)}}_{ab}$, $T^{\text{(lss)}}_{ab}$,
$T^{\text{(cos)}}_{ab}$ representing the matter present at each of
these scales.

Now the Einstein equations may be assumed to hold at the `local' scale: after
all, this is the scale where they have been exquisitely checked, so
\begin{equation}\label{eq:efe1}
R^{\text{(local)}}_{ab} - \frac{1}{2} R^{\text{(local)}} g^{\text{(local)}}_{ab}+ \Lambda
g^{\text{(local)}}_{ab}= \kappa T^{\text{(local)}}_{ab}\,.
\end{equation}
But the averaging process:
\begin{equation}\label{eq:av}
g^{\text{(local)}}_{ab} \rightarrow g^{\text{(gal)}}_{ab} \rightarrow g^{\text{(lss)}}_{ab},\,\,
T^{\text{(local)}}_{ab} \rightarrow T^{\text{(gal)}}_{ab} \rightarrow T^{\text{(lss)}}_{ab},
\end{equation}
doe not commute with evaluating the inverse metric, connection
coefficients, Ricci tensor, and Ricci scalar, for example:
\begin{eqnarray}\label{eq:eeffee}
g^{\text{(local)}}_{ab} \rightarrow g^{\text{(local)}}{}^{ab} \rightarrow
\Gamma^{\text{(local)}}{}^c_{ab} \rightarrow R^{\text{(local)}}_{ab} \rightarrow
R^{\text{(local)}},\\
g^{\text{(gal)}}_{ab} \rightarrow g^{\text{(gal)}}{}^{ab} \rightarrow
\Gamma^{\text{(gal)}}{}^c_{ab} \rightarrow R^{\text{(gal)}}_{ab} \rightarrow R^{\text{(gal)}}.
\end{eqnarray}
Hence if the EFE hold at scale `local' (i.e. equation (\ref{eq:efe1}) is
true), they will not hold at scales `gal' or `lss'; for example one will
find
\begin{equation}\label{eq:efe5}
R^{\text{(gal)}}_{ab} - \frac{1}{2} R^{\text{(gal)}} g^{\text{(gal)}}_{ab}+ \Lambda
g^{\text{(gal)}}_{ab}= \kappa T^{\text{(gal)}}_{ab} + E^{\text{(gal)}}_{ab}
\end{equation}
where the extra term $E^{\text{(gal)}}_{ab} \neq 0$ is due to this
non-commutativity. It is the effective matter source term
representing the effect of averaging out smaller scale structures,
which is then an effective source term for averaged EFE at scale `gal'.
Similarly there will be such an effective such source term
$E^{\text{(lss)}}_{ab}$ at that scale~-- the scale usually represented by
perturbed FLRW models~-- and $E^{\text{(cos)}}_{ab}$ at the cosmological scale. This is the
\emph{backreaction} from the small scales to the larger scales.

In essence, it is an \emph{assumption} that Einstein's equations
also hold for an averaged geometry, as well as a local one. In fact,
it is not clear that the whole machinery of GR holds after
averaging~-- e.g., concepts such as spacetime and objects such as
tensors need to be assumed to make sense after coarse graining.

The classic example of this effect is Isaacson's calculation of the
effective backreaction of small scale gravitational radiation on an
averaged large scale metric \cite{Isa68a,Isa68b}. Now this affect
applies to cosmology: if there is significant gravitational
radiation at early times, this will affect the dynamics at later
times, according to Isaacson's calculation (which can also be
derived from a variational principle, see \cite{MacTau73}). However
that effect will be exceedingly small at late times (it may be
significant at early times, see \cite{AbrBraMuk97,CalHuMaz01}).

We are concerned however with the dynamic effects of non-linear
structure on cosmology at late times (i.e. after decoupling of
matter and radiation). In principle, the same effect will occur in
this context \cite{Ell84};  the question is whether this is a
significant effect or not. We will look at general formalisms and
specific models in the next section.

\subsection{Fitting and observations}

One can do a spacetime fitting, asking which FLRW model is best if
we average invariant quantities in a spatial or spacetime volume:
e.g., the energy density of particles and their velocities (when the
matter averaging may be represented by kinetic theory); one may
choose to smooth the metric or scalar invariants on the geometry
side. Alternatively, one can do a null fitting, where one in effect
averages astronomical observations.

\subsubsection{Fitting and the past null cone}
As we determine our best fit model by null cone observations, this
is in the end what we'd like to do (in effect the standard
observational cosmology approach is this type, but is not usually
phrased this way). So a key issue is how this all relates to
cosmological observations. Here we have to take into account not
only the effect of averaging on the geometry and dynamics (as
represented by (\ref{eq:efe5})) but also the effect of lumps on null
geodesics and on observational relations.

The basic question is: What are observables in these approaches?
What do observations of averaged quantities really tell us about our
Universe? Cosmological observations probe quantities such as the
redshift, the angular diameter distance, the luminosity distance,
and the image distortion, i.e. quantities related to light emitted
by a distance source and propagating on the past null cone.  The
general way this works was described in a perturbative framework in
the pioneering work of Kristian and Sachs \cite{ KandS66}
(extended to more general cases in \cite{EllisMac69,MacEllis70,Elletal85}).
This involves expanding the geodesic deviation vector in orders of
the affine parameter along the past null cone.   Here one relates
the geometric quantities to the observational quantities using the
relation between the redshift and the four-velocity of the observer
$u^a$:
$1+z= \frac{u_ak^a|_e}{u_bk^b|_o}$,
 and the relation between the intrinsic cross-sectional area,
 $dA$ of a distant source and the measured solid angle subtended by the observer
 $d\Omega$:
 $dA= r^2_A d\Omega,$
 where $r_A^2$ is the  angular-diameter distance.
 This expansion as given by  \cite{EllisMac69,MacEllis70} is exactly
 covariant and can be applied to any space time.
 In the present context one would like to compare the general result with that of the FLRW background.
 This is done by matching the background FLRW results with that from the 
 spacetime indicated by observations \cite{Clarksonthesis,Flanagan:2005dk}.

However there are complications. In the real universe, as pointed
out in \cite{Bertotii:1966, KandS66,MacEllis70,Dyer-Roeder},
observations take place via null geodesics lying in the empty
spacetime between galaxies, which are focused only by the curvature
actually inside the beam, not the matter that would be there in a
completely uniform model. The effect on the observational relations
of introducing inhomogeneities into a given background spacetime is
twofold: it alters the redshift, and it changes area distances. This
should also be taken into account in any fitting procedure.

Others have emphasized the importance of smoothing the past
lightcone. A sketch of how averaged Raychauhduri equation on the
past null cone will look like was given in \cite{Coley:2009qc},
while R\"as\"anen in~\cite{Rasanen:2009uw,Rasanen:2008be}  gave the
past null cone averaged equations  for  the scalars in a statistically
homogeneous and isotropic space time. Work in this important area is
still at its infant stage with  no concrete quantifiable physical
result that can be directly fitted to observations.

\subsubsection{Averaging on the past light cone}
Averaging is in some respects a fitting process, but does not
necessarily correspond to any actual observational procedure. Can
one propose an average model of the universe based on the past null cone? A very
recent attempt at a comprehensive approach is Gasperini et
al.~\cite{Gasperini:2011us}.
However this should be approached with caution: observations to
cosmological distances mix spatial and time variation, and it does
not make sense to simply average today's state of the universe with
what it was like in the past. One would expect any averaging
operation to leave the background invariant, and it's not obvious
that this can happen for FLRW somehow averaged on its light cones.
So averaging based on observations would need to involve comparing
the universe today with earlier times by use of dynamical equations
relating variables at these different times: a very model dependent
procedure, and not `averaging' in a normal sense.

What is clear is that backreaction won't be important along the
light cone, because the key causal effects in cosmology propagate in
a timelike fashion \cite{EllSto09}.
Nevertheless it is obviously important to relate the results of
averaging and backreaction effects to observations. The exact and
perturbation approaches that follow attempt to do this.


\section{Non-perturbative backreaction}
\label{NPB}

One  approach is to build a model of the universe `bottom up'. Advocated by
Buchert \cite{Buchert:1999er, Buchert:2001sa, Buchert:2007ik},
Zalaletdinov \cite{Zalaletdinov:1996aj,Zalaletdinov:2008ts},
Wiltshire \cite{Wiltshire:2007jk,Wiltshire:2007fg,Wiltshire:2009ip},
R\"as\"anen \cite{Rasanen:2006kp,Rasanen:2008be,Rasanen:2009uw},
among
others~\cite{1992CQGra...9.1023Z,1992AAS...180.5804Z,1995ApJ...453..574Z,Hoogen:2010qb,Brannlund:2010rs},
these approaches dispute the idea that one needs a background to
work from: rather the background model and its dynamics should
emerge as a large-scale approximation to a more detailed
inhomogeneous model, which can be compared with a FLRW model for the
universe assumed \emph{ab initio} as in the standard approach. As
explained in previous sections, the field equations for such a model
may be expected to be different from those in a standard FLRW model:
we want to understand that difference.

Non-perturbative approaches fall into two main categories. One are
generic averaging formalisms, which aim to understand the nature of
the backreaction terms in general, a bit like deriving and
understanding the macroscopic Maxwell Equations. The other approach
is to create fully relativistic inhomogeneous or even N-body models
by use of simplifying assumptions; then, by comparing observables in
these models with their averaged FLRW counterparts one can hope to
quantify non-perturbatively the backreaction effect and the
magnitude of the fitting problem. Let us consider each approach in
turn, with some of the main attempts in the literature.

\subsection{Averaging Formalisms}

\subsubsection{Buchert's approach}

Alongside early attempts by
\cite{Carfora:1995fj,Russ:1996km,Futamase:1996fk,Boersma:1997yt,Stoeger:1999ig},
Buchert~\cite{Buchert:1999er,Buchert:2001sa} builds on the Newtonian
averaging by Buchert and Ehlers~\cite{Buchert:1995fz,Buchert:1995pj}
to provide a bare-bones approach to the problem, concentrating on
averaging scalars on spatial hypersurfaces. The kinematic scalar
equations for vorticity-free perfect fluid are averaged, to give
evolution equations for the averaged expansion and shear scalars.
Several authors \cite{Buchert:2001sa,Behrend:2007mf,
Larena:2009md,Clarkson:2009hr,Rasanen:2009uw,Gasperini:2009mu} have
generalized this approach to any arbitrary  spacetime, but we will
illustrate with the original Buchert proposal.

Let us assume a dust spacetime, and observers and coordinates at rest with respect to the
dust. The average of a scalar quantity $S$ may be (non-covariantly) defined as simply its integral
over a region of a spatial hypersurface $\mathcal{D}$ of constant
proper time divided by the Riemannian volume:
\begin{equation}
\label{eq:Average}
 \average{S(t,\bm{x})}=\frac{1}{V_\mathcal{D}}{\int_\mathcal{D}
 \sqrt{\det h}\, \d^{3}x \,\, S(t,\bm{x})}
\end{equation}
Taking the time derivative of Eq.~(\ref{eq:Average}) yields the commutation relation
\begin{equation}
\label{eq:CommRel}
[\partial_{t}\cdot,\average{\cdot}]S=\average{\Theta
S}-\average{\Theta}\average{S}\mbox{ ,}
\end{equation}
where $\Theta$ is the expansion of the dust, and we assume the
domain is comoving with the dust. The dimensionless volume scale
factor is defined as $a_\mathcal{D}\propto {V_\mathcal{D}}^{1/3}$,
which ensures $\average{\Theta }=3\partial_t\ln a_D$. Then, the
second derivative of the scale factor is given by the averaged
Raychaudhuri equation:
\begin{equation}\label{Av:Ray}
3\frac{\ddot{a}_{\mathcal{D}}}{a_{\mathcal{D}}}  +4 \pi G
\langle\rho\rangle_{\mathcal{D}}=\Lambda+ \mathcal{Q}_{\mathcal{D}},
\end{equation}
where   $\mathcal{Q}_{\mathcal{D}}= \frac{2}{3}\left[\langle\Theta^2
\rangle_{\mathcal{D}} - \langle \Theta
\rangle_{\mathcal{D}}^2\right]- 2\langle
\sigma^{2}\rangle_{\mathcal{D}} $ is the kinematic backreaction term
and $\sigma^2 =\frac{1}{2}\sigma_{ab} \sigma^{ab}$ is the magnitude
of the shear tensor. The \emph{non-local} variance of the
\emph{local} expansion rate can act in the same way as the
cosmological constant, causing the average expansion rate to speed
up, even if the local expansion rate is slowing down. Even more
tantalisingly, if this were the cause of the observed acceleration,
the coincidence problem would be solved in the most natural way: as
structure forms the variance in the expansion rate grows, as matter
coalesces and virialises~\cite{Rasanen:2003fy}. This is a truly
remarkable possibility in moving from local to non-local quantities
on a non-trivial geometry, and is the reason for the recent
excitement in the averaging problem.

One can see how this counter-intuitive idea works as
follows~\cite{Nambu:2005zn}: If the average scale factor of a
universal domain,  $a_\mathcal{D}$, can be written as  a union of
locally homogeneous and isotropic regions, each with its scale
factor $a_i$, then the  acceleration of the universal domain
$\mathcal{D}$ is given
by~\cite{Moffat:2005ii,Rasanen:2006kp,Wiegand:2010uh}:
\begin{equation}
 a_\mathcal{D}^2\ddot{a}_\mathcal{D} = a_1^2{\ddot a}_1  + a_2^2{\ddot a}_2 + \cdots
                   + \frac{2}{a_\mathcal{D}^3 }
                            \sum_{i\neq j} a_i^3a_j^3
                            \left(
                                  \frac{\dot a_i}{a_i}- \frac{\dot a_j}{a_j}
                            \right)^2 \,,
\label{Toymodel}
\end{equation}
where $a_i$ represents the locally defined scale factor in the
$i$-th sub-region, and, $a_\mathcal{D} \equiv (a_1^3 + a_2^3 +
\cdots)^{1/3}$. Acceleration of the universal domain,  ${\ddot
a}_\mathcal{D} > 0$,  can easily be achieved,  for example,  for a
two disjointed  dust filled FLRW sub-region in which one might be
expanding while the other  is contracting at a time $t$, i.e
$\dot{a}_1 = - \dot{a}_2$ (assuming same sized sub-regions $a \equiv
a_1 = a_2 $), one obtains $
 a_\mathcal{D}^2 {\ddot a}_\mathcal{D}
  =
 2a^3 \left\{\frac{\ddot a}{a} + 4\left(\frac{\dot a}{a}\right)^2 \right\}
  = \frac{7}{3} \kappa^2 a^3 \rho > 0 .
$ Here one easily obtain an acceleration for the universal domain,
${\ddot a}_\mathcal{D} > 0$, even when the two sub-regions are
decelerating  ${\ddot a}_1 < 0$, ${\ddot a}_2 < 0$; i.e  all
observers see only deceleration. However, it has been
argued~\cite{Ishibashi:2005sj}  that acceleration found in this toy
model does not necessary imply that the physical universe is
accelerating, since this model has not been shown to satisfy other
rigorous observational tests. It also ignores problems due to
 matching/junction conditions for any two regions.

In Buchert's scheme, all tensor contributions appear as scalars in
these equations, and are collected into unknown source terms. Of
course, the system of scalar equations is not closed, so one has to
make an ansatz about the effect of averaging the shear terms; so it
is very difficult to say how big the backreaction effect is. One can
derive an evolution equation for the averaged shear scalar; but that
would be sourced by products of Weyl curvature tensors, amongst
other things, and one quickly sees that the system of equations can
never close. So the method of averaging only scalars reaches this
limitation quickly. This feature can further  be understood by
considering the integrability condition:
\begin{equation}
\frac{1}{a_{\mathcal{D}}^6} \partial_t
(\mathcal{Q}_{\mathcal{D}}a_{\mathcal{D}}^6)
+\frac{1}{a^2_{\mathcal{D}}}\partial_t(\average{\mathcal{R}}a_{\mathcal{D}}^2)=0
\end{equation}
where $\average{\mathcal{R}}$ is the average local curvature. This
coupling between the curvature and the volume scale factor implies
that if  $\average{\mathcal{R}} \sim a^{-2}_{\mathcal{D}}$ as
in FLRW cosmology, the kinematic backreaction term will scale as
$\mathcal{Q}_{\mathcal{D}} \sim a_{\mathcal{D}}^{-6}$, which mimics
the behavior of some kind of dark fluid.

Having said that,~\cite{Coley:2009yz} has suggested that, if the scalar curvature invariants can uniquely characterise any
spacetime, a scalar averaging scheme can work in general by
averaging these invariants. He thus arrives at a complete, closed
way of averaging spacetime using only scalars. Whether it is
practical remains to be seen.

The averaged quantities in Buchert's formalism do not
have a clear observational meaning.  Nevertheless, it
is worth noticing that \cite{Rasanen:2008be,Rasanen:2009uw} argues
that in a statistically homogeneous and isotropic Universe, these
average quantities are exactly the ones that describe observations
along the past lightcone. It will be interesting to see if such an argument can be made rigorous.

\subsubsection{Zalaletdinov's Macroscopic Gravity}

A comprehensive approach to covariantly averaging tensors was
initiated and explored by
Zalaletdinov~\cite{Zalaletdinov:1992cf,Zalaletdinov:1992cg,Zalaletdinov:1996aj,Zalaletdinov:2004wd,Zalaletdinov:2008ts} and
later by others \cite{Brannlund:2010rs,vandenHoogen:2009nh,Hoogen:2010qb,Paranjape:2006ww}. It is a
foundational attempt to average the complete set of Cartan structure
equations, in order to define Einstein field equations for the
averaged quantities. This approach is directly inspired by the way
in which a macroscopic theory of electromagnetism can be obtained
from the microscopic Lorentz-Maxwell
theory~\cite{Zalaletdinov:2008ts}. Once averaged, the `macroscopic
field equations' resemble Einstein's, but with a source term~--
analogously to the polarization term in macroscopic Maxwell
equations \cite{Montani:1999}.

The core issue of averaging tensors covariantly is managed by using
bi-local extensions of tensors, so that they transform as tensors at
some point of interest, $x$, but as scalars in a neighbourhood of
the point. Because of that, they can be averaged over that
region~$\Sigma$. The covariant spacetime average is defined
as~\cite{Zalaletdinov:1996aj}
\begin{equation}
\bar{T}_{ab}(x)= \frac{\int_{\Sigma}
\mathcal{A}_a^{a'}(x,x')\mathcal{A}_b^{b'}
(x,x')T_{a'b'}(x')\sqrt{-g(x')}d^4x'}{\int_{\Sigma}\sqrt{-g(x')}d^4x'}
\end{equation} where $ \mathcal{A}_a^{a'}(x,x')$ is the bi-local
transport operator. The backreaction generated by the smoothing
procedure takes the form of a correlation tensor for the
gravitational degrees of freedom which appears as an effective
source in the Einstein field equations. Alternatives to this
definition, and the resulting averaging scheme were initiated
in~\cite{Brannlund:2010rs}.

Implementing this operation results in a completely covariant
smoothing procedure provided the transport operators satisfy certain
conditions. The result is formally independent of the averaging
scale and as such can be seen as generating a universal description
of the collective behaviour of local gravitational degrees of
freedom when only their very large scale properties matter (much the
same way as thermodynamics encompasses the collective behaviour of
particles on very large scales compared to the particles
themselves).

This approach makes an extensive use of transport of tensorial
quantities along geodesics, but by using the natural parallel
transport bitensor the metric is invariant so that no smoothed metric is
obtained. Hence, it relies on a specific choice of bi-tensor that
satisfies a set of differential equations and conditions, in order
to imply the correct properties of the average. This bi-tensor is
used to evaluate integrals on finite domains, and it is not clear
how the formalism is affected by the choice of this
bi-tensor~\cite{Hoogen:2010qb}. Additionally his averaged Einstein
equations rely on some ``splitting rules'' (see eqns (45) and (48)
of \cite{Zalaletdinov:1992cf})
which can be questioned (although they are consistent with an
analysis of high frequency gravitational waves, see eqn (68) of the
same paper).

In~\cite{Coley:2005ei,Coley:2006xu,Coley:2006kp,vandenHoogen:2009nh} it is shown that in
a flat FLRW macroscopic background, the correlation tensor is of the
form of a spatial curvature, while \cite{Paranjape:2007wr} showed
that Zalaletdinov's MG reduces to Buchert's equations with
corrections in an appropriate limit. It has further been employed to
evaluate the backreaction effect in a perturbed FLRW
model~\cite{Paranjape:2008mx,Paranjape:2008ai}, but
the requirement that a FLRW background exists makes this attempt
fall under the category described in Sec.~\ref{FLRW}. While the
amplitude of backreaction is similar to that obtained by simpler
means, the details of using such a covariant approach will be
important for an effective fluid description of perturbations at
second-order, such as~\cite{Baumann:2010tm}.

\subsection{Other approaches}

Another rigorous approach is based on the deformation of the spatial
metric of initial data sets along its Ricci
flow~\cite{Buchert:2002ht,Buchert:2002ij}. In principle, this method
is a nice, natural way of smoothing a space-time that can be linked
to the standard renormalization group approach of effective field
theories~\cite{Carfora:1995fj}, but the non-linearity of the Ricci
flow equations is a serious complication that can lead to the
development of singularities along the flow; that makes its use in a
cosmological context particularly difficult. 
A renormalisation group approach to coarse graining in the very
early universe is given in \cite{CalHuMaz01}.

\subsection{Model building Approaches}

\subsubsection{Timescape Cosmology}
Another very original viewpoint  called  the
Timescape cosmology, has been proposed and investigated by
Wiltshire~\cite{Wiltshire:2007jk,Wiltshire:2008sg,Wiltshire:2009ip}.
It is a brave but contentious attempt to seriously look at the
status of bound regions and their interaction with an
expanding cosmological model.

The idea is to separate the Universe into expanding, underdense regions whose boundaries are
overdense regions enclosing virialized regions 
such as the one we, as observers,
live in. An average is then performed spatially (using Buchert's formalism~\cite{Buchert:2007ik})
to define a reference cosmic background. Interestingly, the amount of backreaction in these models is at most of
order a few percent (when normalized as a fraction of the energy density)
and is not solely responsible for explaining the apparent cosmic
acceleration: the non standard effects principally come from the
desynchronization of local clocks (in the virialized regions) with respect
to cosmic clocks defined via the average background.  Indeed, the gravitational redshift
effects imply different ticking rates for clocks inside the voids and in the virialized regions.
Wiltshire argues that the effect is cumulative when an average is performed to define the background
clocks. A possible interpretation of the model is that the extra redshift effects change the observable
relation in the effective FLRW background. As such, the model is not actually accelerating, but the extra redshift
accounts fully for the dimming of supernovae because they appear to be at
a higher redshift than expected.
A detailed discussion of the observational consequences of the Timescape cosmology can be found in
\cite{Wiltshire:2009db,Leith:2007ay}.

Wiltshire's proposal is more original than simply viewing the problem as one of backreaction of
structures on the overall global dynamics. It also recognises that
the position of the observer (in virialized structures as opposed to
voids) may be important to the fitting problem when the variance in
local geometries becomes large. Nevertheless, it suffers from its
own problems, among which one is the use of a pure two-zones model
to describe the Universe, without proper junction
conditions between the zones (this problem is emphasized in
\cite{Mattsson:2010vq, Mattsson:2010ky}).

\subsubsection{Swiss-Cheese Models}

The Swiss-Cheese model consists of one or more spherically symmetric
vacuum regions, each described by a Schwarzchild metric,
joined across spherical boundaries to a FLRW
model~\cite{EinsteinandStraus,Kantowski1969,Tomita:1999qn,Hellaby:2005ut,Biswas:2007gi}.
This  set-up represents a very natural way  to  model the part of
the universe that we see, for example  
the boundary of a galaxy and intergalactic
space, the lack of effect of the expansion of the universe on the motion of
planets, etc.

The Swiss-cheese type construction can be adopted to study the
effect  of inhomogeneities on cosmological observations in a fully
non-linear and relativistic manner.  The algorithm commonly in use
is to start with FLRW metric  and cut out comoving holes and fill them
with a Schwarzschild metric, while making sure  through the
matching conditions that the  mass in the holes equals the mass that
was removed. Hence, Swiss Cheese models do not affect the
global dynamics of a lumpy universe, i.e there is no dynamical
backreaction effect.

Another more interesting construction is to modify  the homogeneous
FLRW metric  by the introduction of spherical regions of the
spherically symmetric Lema\^itre-Tolman-Bondi (LTB) dust spacetime
\cite{Vanderveld:2006rb,Marra:2007pm, Marra:2007gc,Kolb:2009rp,
Vanderveld:2008vi,Sugiura:1999fm, Clifton:2009nv,
Bolejko:2010eb,Szybka:2010ky}; a quasi-spherical Szekeres model
can also be introduced \cite{Bolejko:2008xh}. One has to ensure
through the matching conditions that the spherical region of
inhomogeneities is comoving and mass compensating,  to ensure that
the LTB and FLRW regions evolve independently. Thus again there are
no dynamical backreaction effects.

There are however significant observational differences from
standard cosmology. These spacetimes model precisely the difference
between Weyl and Ricci focussing of null geodesics. The null
geodesics  in the void regions are focussed only by shear induced by
the Weyl tensor. One of the critical problems in this area is the
relation between Ricci curvature and Weyl curvature, and
Swiss-cheese models are a very interesting way to study this. We
refer to other papers for a discussion of these observational
effects: see \cite{Vanderveld:2006rb,Marra:2007pm, Marra:2007gc,Kolb:2009rp,
Vanderveld:2008vi,Szybka:2010ky,Sugiura:1999fm,Clifton:2009nv} for
differing viewpoints.

\subsubsection{Lindquist-Wheeler type models}
All the preceding models relied on the hypothesis that a
cosmological fluid can be employed to model the distribution of
matter in the Universe. On the contrary, Lindquist-Wheeler type
models are a genuine approach at constructing an expanding universe
model out of locally static domains. It consists in modelling the
Universe by approximately paving its compact spatial sections \footnote{it could  be 
topologically spherical, flat or hyperbolic, but the approximation is better in the case of spherical sections}
(topologically homomorphic to $S^{3}$) 
with Schwarzchild domains
that stand for the static regions constituting the `particles' of
cosmology (such as galaxies)~\cite{LindWheel}. Of course, the
matching cannot be exact, and shells have to be introduced at the
boundaries between the cells. These boundaries then obey equations
of motion that produce an overall expanding and recollapsing model
that closely mimics a $k=1$ FLRW Universe.

This approach fundamentally differs
from a Swiss-Cheese model in that no reference to a FLRW metric is
needed in addition to the static regions: the dynamical properties
really emerge from the interaction between the static cells
encoded in the motion of their boundaries. As such, the fluid
approximation is not required and the fluid-like behaviour only
appears when the dynamics is coarse-grained over the detailed,
local, structure. An interesting point is that it leads to a
solution that is similar to the equivalent FLRW $k=+1$
solution, but different in the details (for example, the relation
between the total mass and the maximum radius is modified).

It has recently been extended interestingly by Clifton and Ferreira
\cite{Clifton:2010fr,Clifton:2009jw} in their Archipelagean cosmology. They showed that the
optical properties of such a model are  different from those
of the "equivalent" FLRW model and can lead to a correction
in the fitted value of $\Omega_{\Lambda}$ of order 10\%. However,
the solution is not, as such, self consistent: it is an
approximation that neglects the interaction between neighbouring
cells, interaction that would result in the deformation of the
geometry around each vertex and the appearance of anisotropies.
But the approximation is plausible and is worth exploring because
it is a neat way to explore the emergence of a collective expanding Universe formed by locally static regions.


\section{Perturbative Backreaction}

All these approaches are enlightening, but do not yet
result in detailed models that can be directly compared with
precision cosmology observations. For that we need to turn to the
standard perturbed FLRW models, which enable us to comprehensively
investigate backreaction effects in the linear and weak non-linear
regime. 

Whether this adequately reflects the dynamics of full
non-linear models can be debated. Paranjape~\cite{Paranjape:2009zu} emphasizes  that the background scale factor, needed to compute the backreaction is affected by the backreaction itself, hence making it impossible to calculate the backreaction precisely as circularity occurs. We assume that structure formation can be described by perturbing a well behaved background (since it seems to give the right power spectrum and indeed everything that has to do with observations of structure formation). The background that has to be fitted to observations, however, is not the one we first thought of with scale factor $a(t)$, but the one with coarse-grained scale factor $a_{D}(t)$. Thus fluctuations change the background that has to be compared with large scale observables, such as the luminosity distances. We return to the issue of the adequacy of these approaches below.

\subsection{Backreaction in the standard model}
\label{FLRW}

The standard model of cosmology ignores all the complexity of
smoothing the spacetime and \emph{assumes} that on `large' scales
(say larger than a few Mpc) we can model the universe as homogeneous
and isotropic, with linear fluctuations describing structure
propagating as smooth fields on this background. On smaller scales
we can jump to Newtonian gravity, and model the universe as discrete
particles in simulations. Because it is the only model we have where
we can calculate anything realistically at all, it is the perfect
arena to study backreaction in detail. We shall give a rough
overview of the issues involved, which illustrate more about the
backreaction problem in cosmology.

The fields that are propagating on the background alter its dynamics
through the non-linearity of the field equations. Could the gravity
of the gravitational potential be important? According to observers
at rest with the gravitational field, the potential itself is small
everywhere outside objects less dense than neutron stars, and so if
we write $g=g_0+h$ where $h\ll1$, how might the backreaction of this
perturbation $h$ add up to something large? The affine connection or
Ricci rotation coefficients determine the dynamics of the spacetime,
and are generically $\mathcal{O}(\partial h)$; the field equations
are $\mathcal{O}(\partial^2 h)$. A perturbation of wavelength
$\lambda= a/k$, where $k$ is the comoving wavenumber and $a$ is the
scale factor, much less than the Hubble scale can give rise to large
fluctuations in the field equations $\mathcal{O}(k^2 h)$, even
though the change from the background metric is small. Such terms
describe density fluctuations, and  these can be large even though
the metric potentials are small. Can this change the background
significantly?

Let us consider this in some detail, in the simplest cosmology which
agrees with observations: a flat LCDM model with Gaussian scalar
perturbations. Averaging FLRW perturbations has been discussed in
different guises in the literature (mostly for an Einstein-de Sitter
model): some authors investigate specifically the modification to
the Hubble expansion rate or other
variables~\cite{1995ApJ...453..574Z,Russ:1996km,Boersma:1997yt,Rasanen:2003fy,Kolb:2004am,Li:2007ci,Li:2007ny,Li:2008yj,Kolb:2009rp,Clarkson:2009hr,Rasanen:2010wz,Umeh:2010pr,Clarkson:2011uk};
others reformulate the average of the backreaction into an effective
fluid~\cite{Noonan1984,Noonan1985,Paranjape:2006ww,Paranjape:2007wr,Paranjape:2007uj,Behrend:2007mf,Brown:2008ra,Peebles:2009hw,Baumann:2010tm,Chung:2010xx},
while one of  the first attempts considered the important problem of
how to calculate the averaged metric~\cite{Stoeger:1999ig}. Rather
than summarize these approaches, let us discuss what happens in a
general way.

In the Poisson gauge to second-order in scalar perturbations the
metric reads~\cite{Bartolo:2005kv,Baumann:2010tm} \bea \d
s^2&=&-\left(1+2\Phi+\Phi^{(2)}\right)\d t^2 + 2 V_i \d x^i \d
t \nonumber\\&& +a^2\left[\left(1-2\Psi-\Psi^{(2)}\right)\delta_{ij}
+ h_{ij}\right] \d x^i \d x^j, \label{metric:Newton-FLRW} \eea
where $\partial_iV^i = 0$, $h^i_{\,\, i} = 0$,
$\partial_ih^{ij} = 0$ (because scalar, vector, and tensor modes
interact at second order, it is inconsistent to include scalar modes
alone). The background evolution of the scale factor $a(t)$ at late
times is determined by the Friedmann equation \be
H(a)^2=\left(\frac{\dot
a}{a}\right)^2=H_0^2\left[\Omega_ma^{-3}+1-\Omega_m\right] \ee where
the Hubble constant $H_0$ is the present day expansion rate, and
$\Omega_m$ the normalised matter content today. The first-order
scalar perturbations are given by $\Phi, \Psi$ (and are all that is
required for observations at the moment), and the second-order by
$\Phi^{(2)}, \Psi^{(2)}$ (which are needed for a consistent analysis
of backreaction). In this gauge we have the metric in its
Newtonian-like form, which we may think of as the local rest-frame
of the gravitational field because it is the frame in which the
magnetic part of the Weyl tensor vanishes for vanishing vector and tensor perturbations~\cite{Clarkson:2009hr}.

For a single fluid with zero pressure and no anisotropic stress
$\Psi=\Phi$, and $\Phi$ obeys the `master' equation
\begin{equation}
\ddot\Phi+4H\dot\Phi+\Lambda\Phi=0\,.
\label{eq:BE}
\end{equation}

For a LCDM universe the solution in time to this is equation has
$\Phi$ constant until $\Lambda$ becomes important, and then starts
to decay as $\Lambda$ suppresses the growth of structure on all
scales by about a factor of 2. There is no scale dependence in the
equation, which all comes from the initial conditions~-- usually a
nearly scale-invariant Gaussian spectrum from frozen quantum
fluctuations during inflation~-- and subsequent evolution during the
radiation era. In Fourier space, assuming scale invariant initial
conditions from inflation, the power spectrum of $\Phi$, ${\cal
P}_\Phi$, is independent of scale for modes larger than the equality
scale, $k_{eq}\approx0.07\Omega_mh^2$Mpc$^{-1}$, and $\sim
k^{-4}(\ln k)^2$ for modes much smaller than it, up to some
non-linear scale $k_{NL}\ll k_{eq}$. The change in behaviour at the
equality scale, arising from modes which enter the Hubble radius
before matter-radiation equality, is important for backreaction
because it is the modes larger than the equality scale which are
primarily responsible for any backreaction at all. In essence, the
equality scale determines the size of the backreaction effect.

All first-order quantities can be derived from $\Phi$; for
example,
\be\label{vel}
v^{(1)}_i=-\frac{2}{3a^2H^2\Omega_m}\partial_i\left(\dot\Phi+H\Phi\right),
\ee
is the first-order velocity perturbation, which governs the peculiar
velocity between the matter flow and the rest-frame of the
gravitational field. Meanwhile, the gauge-invariant density perturbation is
\be
\delta=\frac{\delta\rho}{\rho}=\frac{2}{3H^2\Omega_m}\left[a^{-2}\partial^2\Phi-3H\left(\dot\Phi+H\Phi\right)\right]\,.
\ee
The second-order solutions for $\Psi^{(2)}$ and $\Phi^{(2)}$ are
given by~\cite{Bartolo:2005kv}. These are complicated expressions involving
time integrals over products of $\Phi$ and its derivatives. For
backreaction, however, the important thing is that these
contain terms of the form $\partial_i\Phi\,\partial^i\Phi$, as
well as non-local terms such as
$\partial^{-2}(\partial_i\Phi\,\partial^i\Phi)$.

What is the backreaction of perturbations onto the the expansion rate?
At second-order, after substituting for $v^{(1)}_i$ from Eq.~(\ref{vel}), the expansion rate looks something like~\cite{Rasanen:2003fy,Kolb:2004am,Li:2007ci,Li:2007ny,Clarkson:2009hr,Rasanen:2010wz,Umeh:2010pr,Paranjape:2006ww,Paranjape:2007wr,Paranjape:2007uj,Behrend:2007mf,Brown:2008ra,Clarkson:2011uk}:
\bea
{\cal H}&=&H+\text{1st-order terms like}~\Phi~\text{and}~\partial^2\Phi \nonumber\\
&&+\text{2nd-order terms like}~\Phi^2,~\Phi\partial^2\Phi,~\Phi_2,~\Psi_2,~\partial_i v_2^i\nonumber\\
&&\text{and time derivatives thereof}. \eea 
Provided the domain is
small, this quantity can also correspond to the sky-averaged Hubble
rate, measured from observations on the past
nullcone~\cite{Barausse:2005nf,Clarkson:2011uk}. In principle, then,
the perturbative terms give the backreaction to the local expansion
rate. To evaluate $\cal H$ we can use a realisation of $\Phi$ given
an inflationary model.  Alternatively, we can assume a spectrum for
$\Phi$ and evaluate the statistics of $\cal H$. This allows us to
calculate the expectation value of $\cal H$, $\overline{\cal H}$, as
well as its variance, in terms of integrals over the power spectrum
of $\Phi$ multiplied by powers of $k$. The reason we must go to
second-order now becomes clear when we calculate the expectation
value: for Gaussian perturbations from inflation, the ensemble
average of $\Phi$ is zero, which implies~-- assuming ergodicity~--
that when averaged on the background over a large (strictly,
infinite) domain they are zero too. Thus, the second-order terms
provide the main backreaction effect; the first-order terms give the
variance.

\subsubsection{Averages used}

Before going further, it is worth summarizing the different types of
averaging that are used in this kind of analysis:
\begin{description}
\item[Riemannian averaging] this is the `correct' way
to average scalars in a non-homogeneous and non-isotropic spacetime.
In perturbation theory it can be expanded in terms of the Euclidean
average, introducing products of averaged quantities. It doesn't
commute with the time derivative, implying more non-connected terms.
Acting on a second-order quantity, this is the same as Euclidean
averaging at that order.
\item[Euclidean averaging or Smoothing] This is spatial averaging on the background.
When thinking of perturbations in the metric as fields on the
background this is the natural averaging to use. When thinking of
the perturbed metric as an approximate spacetime in its own right,
quantities should be smoothed intrinsically using the Riemannian
averaging.  Smoothing is the same, but often applied directly to
$\Phi$ to remove structure below a given scale. It was used
in~\cite{Baumann:2010tm} to integrate out small scales, where it was
applied to tensor components as well as scalars.
\item[Ensemble averaging] The key tool for statistical evaluation, assuming Gaussian
perturbations from inflation. A Euclidean average over an infinite domain is the same
as ensemble averaging, assuming ergodicity.
\end{description}

It is often assumed that these are interchangeable, but they are
not. In particular, the scale dependence of averaging is very
different depending on the procedure used, and this could have an
impact observationally~\cite{Li:2007ci,
Clarkson:2009hr,Umeh:2010pr}. Furthermore, Euclidean averaging~-- or
often even Riemannian~-- is often replaced with an ensemble average,
loosing the non-connected terms which are responsible for
scale-dependence. This gives only the super-Hubble contribution to
backreaction. The exact relation between ensemble averaging and
Riemanning averaging is another open important question.

\subsubsection{Averaging}

Consider now the average of $\cal H$ over a spatial domain $\cal D$,
where the spatial surfaces are defined in the coordinate frame. The
Riemannian average of a quantity $\Upsilon$, 
\be \label{av}
\langle
\Upsilon\rangle_{\CD}=\frac{1}{V_{\CD}}\int_\CD \sqrt{\det h}\, \d^3
x \Upsilon 
\ee
 can be expanded in terms of the Euclidean average  defined on the background space slices,
$\langle \Upsilon \rangle ={\displaystyle\int_\CD\d^3x\, \Upsilon }\bigg/{\displaystyle\int_\CD \d^{3}x}$, as:
\begin{equation}
\label{ExpandAverage} \average{\Upsilon}=
\Upsilon^{(0)}+\langle\Upsilon^{(1)}\rangle+\langle\Upsilon^{(2)}\rangle+
3\left[\langle\Upsilon^{(1)}\rangle\langle\Psi\rangle-\langle\Upsilon^{(1)}\Psi\rangle\right]\mbox{,}
\end{equation}
where $\Upsilon^{(0)}$, $\Upsilon^{(1)}$ and $\Upsilon^{(2)}$ denote
respectively the background, first order and second order parts of
the scalar function
$\Upsilon=\Upsilon^{(0)}+\Upsilon^{(1)}+\Upsilon^{(2)}$. Note the
important term in square brackets, which encapsulates the
relativistic part of the averaging procedure. Distinct types of
terms now appear in the averaged expansion rate: \be \average{\cal
H}\sim\left\{\begin{array}{lr|l}
 \text{1st-order terms} &  \saverage{\Phi} &  \saverage{\partial^2\Phi} \\
 \text{2nd-order `connected'} &  \saverage{\Phi^2} & \saverage{\Phi\partial^2\Phi}  \\
 \text{2nd-order `non-connected'}   &  \saverage{\Phi}\saverage{\Phi}   &   \saverage{\Phi}\saverage{\partial^2\Phi}
\end{array}\right.
\ee
This smoothed Hubble parameter may be the closest variable to
what is measured in practise~\cite{Li:2007ci,Li:2007ny,Li:2008yj},
at least for small domains where the linear Hubble law applies.
Again, the expectation value (ensemble average) and variance can be
calculated using the statistics of $\Phi$. Once the ensemble average
is taken, the first-order terms drop out, the 2nd-order connected
terms become independent of scale (as is the case  for all terms in
$\overline{\cal H}$), while only the non-connected terms retain
scale dependence. Ensemble averages of pure divergence terms, such
as $\partial_i v^i_2$ (which actually contains $(\partial^2\Phi)^2$
terms~-- see below) also drop out~\cite{Clarkson:2009hr}; these are
usually dropped in backreaction studies because they are boundary
terms, so by statistical homogeneity must be
small~\cite{Rasanen:2010wz}.

A similar analysis has been carried out for average of the
deceleration parameter $q=-(1+\dot{H}/H^2)$, which quantifies the
rate of change of the local Hubble rate; it can be extended by
replacing $H\mapsto\saverage{\mathcal{H}}_\CD$ to measure the
deceleration of the average Hubble rate. These are different things,
and have either $\saverage{\partial^2\Phi\partial^2\Phi}$ or
$\saverage{\partial^2\Phi}\saverage{\partial^2\Phi}$ terms. Such
terms also appear in the deceleration parameter defined via a series
expansion of the distance-redshift relation, which is a physical
observable~\cite{Barausse:2005nf,Clarkson:2011uk}.

The relations for determining the scaling behaviour for the
backreaction terms are \bea\label{av-phi}
\overline{\saverage{\partial^m\Phi\partial^n\Phi}}
\sim\int_{0}^\infty \d k\, k^{m+n-1}\mathcal{P}_\Phi(k),\nonumber\\
\overline{\saverage{\partial^m\Phi}\saverage{\partial^n\Phi}}
\sim\int_{0}^\infty \d k\,
k^{m+n-1}W(kR_D)^2\mathcal{P}_\Phi(k),\nonumber \eea
where $W$ is an
appropriate window function specifying the domain. Note that the
connected terms have no dependence on the domain size at
all. Given the approximate behaviour of ${\cal
P}_\Phi\sim\Delta_{\cal R}^2\sim10^{-9}$ for $k\ll k_{eq}$, and
${\cal P}_\Phi\sim\Delta_{\cal R}^2(k_{eq}/k)^4\ln(k/k_{eq})^2$ for
$k\gg k_{eq}$, the scaling of the backreaction terms may be
estimated. For this we need to replace
$\int_{0}^\infty\mapsto\int_{k_{IR}}^{k_{UV}}$. Then, the key
scalings for backreaction become: \bea
\overline{\saverage{\Phi^2}}&\sim& \Delta_{\cal R}^2\ln\frac{k_{eq}}{k_{IR}}\nonumber\\
\overline{\saverage{\Phi\partial^2\Phi}}&\sim&  \Delta_{\cal R}^2 k_{eq}^2\nonumber
\eea
Terms like these appear in the average of the Hubble rate, as well as its ensemble average.

The second type of term,
$\overline{\saverage{\Phi\partial^2\Phi}}\sim k_{eq}^2$, arising
from peculiar velocity terms, is the term primarily responsible for
setting the fundamental amplitude of the backreaction in the Hubble
rate. It is quite small, \be
\overline{\saverage{\Phi\partial^2\Phi}}/(\Omega_mH_0^2)\sim
\Delta_{\cal R}^2 k_{eq}^2/\Omega_m k_H^2\sim  \Delta_{\cal R}^2
T_{eq}/T_0\sim 10^{-5} \ee
 for the concordance model. (The overall
effect is somewhat larger than this due to the contribution of
several such terms.) So, we're looking at sub-percent changes to the
Hubble rate from backreaction, though non-connected terms make it
larger on small scales. Yet, we can observe that the
\emph{backreaction is small because the equality scale is large in
our universe}, which is because the temperature of matter-radiation
equality is very low. Modes which enter the Hubble radius during the
radiation era are significantly damped compared to those which
remain outside until after equality; so, the longer the radiation
era, the less power there is on small scales to cause a significant
backreaction effect.
For a scale-invariant spectrum, then, it may be considered that the
long-lived radiation era is the reason that the dynamical
backreaction is small. The temperature will have to drop by several
orders-of-magnitude before backreaction in the Hubble rate becomes
significant.

The first type of term, $\Phi^2$, is nominally much
smaller,~$\mathcal{O}(10^{-10})$. However, it also tells us that the
IR divergence in $\Phi^2$ must be cut-off by hand. This is
interesting because it implies that a scale invariant spectrum can
not go on forever, and so in this sense the universe cannot be
infinite. This implies that backreaction could in principle be used
to place limits on the start on inflation, which governs the largest
mode $k_{IR}$, which is the first mode to leave the Hubble radius
during
inflation~\cite{Kolb:2004am,Kolb:2005me,Barausse:2005nf,Noh:2009yu}.
There has been speculation that this might lead to very important
effects and even mimic dark
energy~\cite{Kolb:2004am,Kolb:2005da,Barausse:2005nf}, though this
has been
criticised~\cite{Flanagan:2005dk,Hirata:2005ei,Geshnizjani:2005ce}.
However, for this to be significant we require $k_{IR}/k_{eq}\sim
10^{\sim10^{4}}$ (where $\Phi^2$ would compete with the other
terms), which is quite large.  Alternatively, given that we only
measure the primordial power spectrum to be nearly scale-invariant
over a comparatively narrow range of scales, the appearance of
$\overline{\saverage{\Phi^2}}$ implies that it can't be too tilted
to the red on super-Hubble scales. A red spectrum would convert the
logarithmic divergence into a power-law one, and constraints on the
largest mode would be much stronger (e.g., $k_{IR}/k_{H}\lesssim
10^{\sim80}$ for $n_s=0.95$).

In both the variance of the Hubble rate, and the backreaction on
$q$, much larger terms appear: ${\partial^2\Phi}{\partial^2\Phi}$,
which are of order the density fluctuation squared. (While they
appear in the Hubble rate through the second-order velocity
perturbation, the ensemble average conspires to cancel them out.)
This now behaves as,
$\overline{{\partial^2\Phi}{\partial^2\Phi}}\sim \Delta_{\cal R}^2
k_{eq}^4\times[\text{divergent integral}]$, and so: \bea
\frac{\overline{\saverage{\partial^2\Phi\partial^2\Phi}}}{(\Omega_mH_0^2)^2}&\sim&
\frac{\Delta_{\cal R}^2}{\Omega_m^2} \frac{k_{eq}^4}{ k_H^4} F\left(\frac{k_{UV}}{k_{eq}}\right)
\nonumber\\
& \sim &  \underbrace{\Delta_{\cal R}^2 \left(\frac{T_{eq}}{T_0}\right)^2}_{\displaystyle\sim 10^{-2}}
F\left(\frac{k_{UV}}{k_{eq}}\right)
\eea 
which is pretty significant in size, and diverges as
$k_{UV}\to\infty$. The function $F$ is roughly $F(x)\sim 0.5
x^{2.1}$ for $1\lesssim x\lesssim 10$, $\sim 70
x^{-0.1}(\log_{10}x)^{4.75}$ for $x\gg1$, and approaches $\sim
53\ln^3 x$ as $x\to\infty$. Importantly, $F$ overcomes the
pre-factor around $k_{UV}\sim 10k_{eq}$, so these terms are big, and
difficult to know what to do with: here, the UV cutoff is really a
measure of our ignorance. Within linear perturbation theory it
should be set by the end of inflation and the reheating temperature,
as well as the small-scale physics of dark matter, both of which are
sub-pc scales today. Replacing the UV cutoff with a smoothing
function in $\Phi$ implies that we might do better to smooth
order-by-order, and calculate second-order terms from smoothed
first-order ones, rather than average directly at a given order.
Even for domains much larger than the non-linear scale, where linear
perturbation theory breaks down (somewhere around a few Mpc), $F$ is
quite sizeable, and so we have backreaction terms ${\cal O}(1)$.
From this, we also recover that the variance in the Hubble rate is
${\cal O}(1)$ on scales of Mpc.

The divergencies arise not only because of the high derivative terms
which appear at second-order but also because of the scale invariant
initial conditions and power-law suppression of modes below the
equality scale. Without this suppression, even the $\Phi^2$ terms
would lead to significant backreaction, as modes all the way up to
the inflationary cutoff would contribute. In such a scenario it is
clear that the canonical approach to perturbation theory would fail
completely. As we discuss below, a reformulation of perturbation theory might avoid these type of problems~\cite{Baumann:2010tm}.

\begin{figure}[htbp]
\begin{center}
\includegraphics[width=0.8\columnwidth]{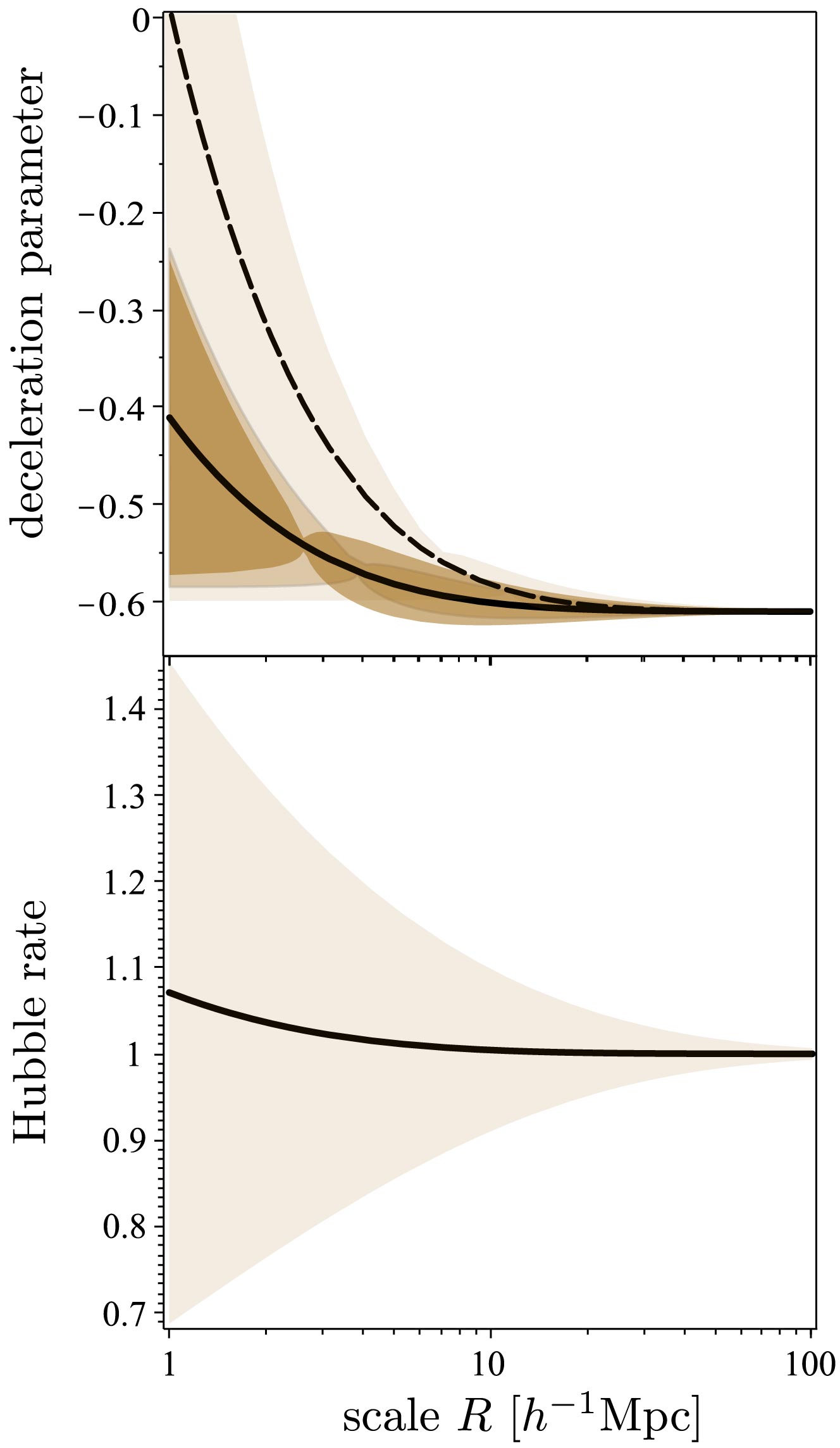}
\caption{The averaged Hubble rate$/H_0$ and the deceleration parameter, after
averaging, as a function of scale, for the concordance model with
$\Omega_m=0.26, f_b=0.17, h=0.7,
\Delta_\mathcal{R}^2=2.49\times10^{-9}$. The heavy black lines shows
the deviation from the background, and the shaded region the
ensemble variance as a function of averaging scale (that is, it is
the variance one gets randomly throwing down Gaussian averaging balls of
radius $R$). For the deceleration parameter there is some subtlety in the
definition~\cite{Clarkson:2011uk}, which give different variances and
scales: from outside to in we have the deceleration of the averaged
scale factor (Buchert's definition, dashed), the spatially averaged deceleration parameter, and
the ensemble average of the local deceleration parameter, now as a
function of the UV cutoff (here evaluated using a smoothing function
on $\Phi$). The UV cutoff completely dominates the amplitude.\label{figure}} 
\end{center}
\end{figure}

Would higher-order perturbation theory affect these conclusions? On
the one hand it seems clear that higher-order perturbations should
be suppressed. Provided $\Phi$ is Gaussian, only even orders will be
important, once ensemble averages are taken. We might expect the
largest terms at any order $n$ to behave like
$\Phi^{(n)}\sim(\partial\Phi)^n$ (e.g., from relativistic
corrections to the peculiar velocity), the ensemble average of which
goes like $\Delta_{\cal R}^n k_{eq}^n$. Terms which appear in the
Hubble rate at order $n$ of the form $\partial^2\Phi^{(n)}$ do not
have enough derivatives to overcome the suppression from
$(\partial\Phi)^{(n-2)}$ terms. By this argument, second-order
should be as large as it gets, and backreaction from structure is
irrelevant. On the other hand,
others~\cite{Notari:2005xk,Rasanen:2006kp,Rasanen:2010wz} have
argued that at higher order terms such as
$(\partial\Phi)^2(\partial^2\Phi)^{n-2}$ are the norm; in this case,
from 4th order on perturbation theory formally diverges~-- at least
as far as calculating averages is concerned. Even if not divergent,
if $(\partial^2\Phi)^{n-2}\sim 1$ then higher-order terms are at
least as large as at second-order and must be included to evaluate
backreaction properly, and so correctly identify the
background~\cite{Clarkson:2011uk}. But do these terms cancel out?
This is definitely an open important problem. It is likely that an
extension of the successful methods of~\cite{Crocce:2005xy}, which
have worked so well for Newtonian perturbation theory, will be
important, but they must be extended to include vector and tensor
degrees of freedom~-- a significant challenge.

It is intriguing that the amplitude of backreaction can be dominated
by the UV cutoff, and so difficult to quantify. In some ways this
seems rather unphysical: we know that spherical systems cannot alter
the expansion rate, which follows from Birkhoff's theorem. Recent
work by Kolb~\cite{Kolb:2011} argues that this in fact sets the cutoff
and so backreaction is dominated by that scale. Alternate work by
Baumann \emph{et al}~\cite{Baumann:2010tm}, which we discuss further
in a following subsection, has argued that when a system is smoothed
over a scale larger than its size the virial theorem holds, in a
pseudo-Newtonian limit. This holds in the Newtonian case where
sources can be thought of as `localised', in equilibrium, and so
freeze out of the cosmic expansion. Of course, such a general theorem cannot
exist in general relativity because energy is radiated to infinity,
and only the stationary part of a system
virialises~\cite{Gourgoulhon:1994}. Nevertheless, in the same way
as for spherical structures this suggests the Newtonian
virial scale is reasonable for the cutoff, as scales below this are shielded from the expansion. However, even a conservative cut-off at
a few Mpc gives backreaction effects of order unity (see Fig.~\ref{figure}), the precise
amplitude of which is completely dominated by how the cutoff is
performed (a UV cutoff, vs a formal smoothing procedure, say). It is
an open problem as to how this should be achieved in practise in a
precise way: the cutoff is there to provide an ad hoc adjustment to
the \emph{linear} spectrum to simulate the \emph{non-linear}, which has not been calculated. 

One can argue that such arguments are in fact irrelevant. As a referee has commented to us: ``As voids
form they occupy a larger and larger fraction of the volume, since they
expand faster than the mean expansion. The increase with time of the
fraction of volume in which the expansion is faster than average means a
decreasing average deceleration, and with smaller smoothing volumes
fully developed voids can occupy larger fractions of the smoothing
volume, increasing the variance. However, voids cannot expand faster
than the average expansion indefinitely, since they collide with
neighboring voids. Calculations [\ldots] which fail
to take void collisions into account, cannot hope to give a correct
account of the large-scale average deceleration once voids on any scale
are fully developed. Since most of the matter ends up in the walls,
filaments, and clusters between voids, which contain almost all the
galaxies, what is relevant for the observed average deceleration on
large scales is the deceleration of these walls and filaments with
respect to each other, which has nothing to do with a volume-averaged
expansion rate dominated by void interiors. The large scale deceleration
is not a question which can be addressed by perturbation theory to the
extent that backreaction coupling small-scale nonlinearities to the
large scale dynamics has any importance." Though it is not clear how one envisages the wall/void interaction, the void dynamics will clearly be important. This alternative view highlights the kinds of interesting issues that remain unresolved.

We now discuss two recent approaches which may provide motivation around some of the problems we have discussed.

\subsection{Short Wavelength  Approximation}

The problem associated with the ill-definition of an average for tensors was
examined by Green and  Wald  in~\cite{Green:2010qy}, generalizing
earlier work by Burnett \cite{Burnette:1988}.
They replaced averages by the notion of weak limits in the weak field
approximation, and thereby obtained strong restrictions on
back-reaction effects (in this context) through a mathematically
precise point limit process. In this approximation, they show that
if the small-scale motions of matter inhomogeneities are
non-relativistic, the effect of small scale inhomogeneities on large scale dynamics
can be  written as an  effective trace-free (radiation) stress
energy tensor, and hence cannot lead to acceleration via a negative
active gravitational mass.

However the degree to which this analysis captures the physically
relevant degrees of freedom is debatable  because of the nature of
the ultra-local limiting process, which ignores all details of the
actual clustering of matter; but we shall see below that whether
backreaction effects are significant or not depends crucially on the
nature of that clustering. Furthermore,
this analysis is based on the work in \cite{Burnette:1988}, which centers on
handling a singular limit at the origin (a vanishing energy momentum
tensor at all points away from the origin has a non-zero limit at
the origin) arising through behaviour of the `$x \sin(1/x)$'
variety. But  does that realistically represent any real physical
matter distribution in cosmology? Indeed it seems likely that the
crucial quantity representing this discontinuity ($\mu_{mnabcs}$ in
\cite{Burnette:1988}, $\mu_{abcdef}$ in \cite{Green:2010qy}) will vanish for
any realistic matter distribution: does the kind of ultra-local
backreaction mechanism envisaged in these papers  occur in
physical reality? It is not generically the same as the backreaction
due to averaging over finite volumes that is the concern of this
paper, although it might be a limit of such a mechanism in specific
singular geometric circumstances.

\subsection{Effective Fluid Approach}

In order to avoid the divergence problems caused when $\delta$
exceeds unity,~\cite{Baumann:2010tm} have
considered a reorganization of the perturbative expansion, using a
coordinate based Euclidean smoothing to separate long and short
wavelength modes.  Further, as we have discussed,
$v^2\sim(\partial\Phi)^2\sim\Phi$ in magnitude, and on small scales
when $\delta\sim1$ a natural expansion variable is $v^2$, provided
each spatial derivative reduces the order by $v$. The field
equations are linear in the matter variables, so there is no need to
expand $\delta$. Using this, averaging over suitable scales, one can
derive effective pressure and densities which, when averaged, obey
Newtonian-like equations~\cite{Peebles:2009hw} for the kinetic and
potential energies on small scales and proving a viral theorem for
local systems imbedded in the expanding universe. From this they
argue that backreaction is always small, even in a model with no
radiation era, and (in agreement with Green and Wald) cannot
generate a negative active gravitational mass ($\rho +3p \geq 0$
always). In effect, the potential for a negative gravitational mass
to be generated by Buchert's averaging process is vitiated because
of the existence of local equilibrium states characterised by their
virial theorem.

The main idea behind the effective fluid approach proposed in
\cite{Baumann:2010tm} is to re-write the Einstein equations into the
background,  forms linear in $X$, and those non-linear in $X$:
\begin{equation}
\bar G_{a b}+ (G_{a b})^{\rm L}[ X] + (G_{a b})^{\rm NL}[ X^2] =  T_{a b}\, .
\end{equation}
and to assume that the background equations, $\bar G_{a b} =  \bar
T_{a b}$, and the linearized Einstein equations, $(G_{a b})^{\rm L}
= (T_{a b})^{\rm L}$, are  defined in the standard way. Then the
Einstein equations may be  written in a form that is very similar to
the linear equations,
$(G_{a b})^{\rm L} = \, (\tau_{a b} - \bar T_{a b}),$
where the  effective stress-energy pseudo-tensor $\tau_{a b} $ may
then be defined as,
$\tau_{a b}  \equiv  T_{a b} - (G_{a b})^{\rm NL}.$ The second part
in this process requires that the perturbation on the right hand
side of the field equation be performed in orders of the peculiar
velocity instead of density, since the density contrast is
ill-defined at non-linear scales. 
Then at some scale $\Lambda^{-1}$
each linear term is  split into short wavelength modes and the long
wavelength modes as, $ X = X_\ell + X_s ,$ and the non-linear splits as,
\begin{equation}
 \langle fg\rangle_\Lambda = f_\ell g_\ell + \langle f_s g_s\rangle_\Lambda +
 \frac1{\Lambda^2}\nabla f_\ell \cdot \nabla g_\ell + \dots \; .
\end{equation}
After smoothing, the effective energy momentum pseudo tensor becomes,
\beq
\label{equ:effTau}
\saverage{\tau_{ab}}_\Lambda =\saverage{\tau_{ab}}^\ell + \saverage{\tau_{ab}}^s +  \saverage{\tau_{ab}}^{\partial^2} \,.
\eeq
The superscripts $s$, $\ell$ and $\partial^2$ denote the short wavelength,
the long wavelength and suppressed higher derivative parts, and $\Lambda $ is the cut-off
for the effective theory. The tensor $\tau^a{}_{b}$ is conserved by
virtue of the linearized Bianchi identity, and  can be re-written
into the form of a fluid with density, pressure and anisotropic
stress,
\begin{eqnarray}
&&\rho_{\rm eff}= \overline{ \saverage{\tau_{ab}}^s}  \tilde{u}^a_\ell \tilde{u}^b_\ell ,\  3 p_{\rm eff} = \overline{\saverage{\tau_{ab}}^s}\gamma^{ab}_\ell ,\nonumber\\ &&\Sigma_{\langle ij\rangle}^{\rm eff}
\approx \overline{\tau}_{\langle i j\rangle}  ,
\end{eqnarray}
where $\tilde{u}_\ell^a$ is the renormalized matter 4-velocity and
overline denotes ensemble average.
Non-linear terms in
$\tau_{a b}$ may be re-written in the form of the kinetic energy
$\kappa $ and the potential energy $\omega$, which evolve as $d(\kappa+\omega)/dt+H(2\kappa+\omega)=0$. The effective density and pressure are given by $\rho_{\rm eff}=\bar \rho_m (1+\kappa + \omega)$ and
$ 3 \bar p_{\rm eff}  = \bar \rho_m (2\kappa + \omega)\, , $ and its
equation of state becomes $ \label{equ:eos} \bar w_{\rm eff} \equiv
\frac{\bar p_{\rm eff}}{\bar \rho_{\rm eff}} = \frac{1}{3} (2\kappa
+ \omega)\,. $  
They argue that structure below the virial scale decouples from the effective
long wavelength expansion of the universe and effective pressure
vanishes: $ 2\kappa + \omega=0$.  
The more detailed
version of this proof in \cite{Baumann:2010tm} relies on the fact that within the
sub-horizon region, one can safely ignore the expansion of the
universe (so one can set $a=1$) and that the smoothing domain is
much larger than the size of the system. Setting $a=1$ and stating that a system is localised is equivalent
to imposing a stationary orbit condition as was done in
\cite{Gourgoulhon:1994} for a general relativistic version of the
virial theorem.

The effective theory they develop holds on large scales $k\ll\Lambda$ and contains a set of effective fluid 
parameters which must be determined from N-body simulations or directly from observations, or from higher-order perturbation theory. 

This paper is a significant and interesting study of the back
reaction issue, calculating some of the effects of averaging and
taking the relevant scales into account, to see when it may be
important. It brings together ideas of both Buchert and Zalaletdinov of deriving effective field equations and hence an effective fluid which hold on macroscopic scales.  They also take care to motivate how virialised regions are cut off from the global expansion, as Wiltshire has emphasised. They approach the problem from the point of view of the standard model, allowing more quantitative predictions, but consequently  suffer from relying on the assumptions of the existence of a global background geometry. 

One of the major differences between \cite{Baumann:2010tm} and other works lies in the fact that they are unconcerned with the problems of \emph{relativistic} averaging, which is emphasised as critical by many authors discussed in Sec.~\ref{NPB}. All their smoothing is performed on the background, as all fields are considered as fields on that background. Averages of tensors are performed on tensor components. We can see from the general definition of the Riemannian average of a scalar, Eq.~(\ref{av}), that if Riemannian averaging were used instead extra terms such as $\Phi\delta\sim v^2\delta$ would appear in their effective fluid which could change the nature of the result (though probably not the overall amplitude). Would this require just a re-definition of the effective fluid parameters, or something more significant?  Related issues are that their results rely on working in a ``good" gauge, and the assumption that coordinates exist which cover the whole spacetime on both small and large scales adequately; this is of course in common with most perturbative approaches.

\subsection{Discussion}

What can we conclude from this? One could conclude that
dynamical backreaction is small by good fortune: the universe is so
hot and had such a long radiation era that small-scale power is
significantly reduced over its scale-invariant initial conditions.
The backreaction terms in the
Hubble expansion, $\Phi\partial^2\Phi$, are the largest ones
which appear in the lhs of the Einstein Field Equations, because the
Einstein tensor has at most two derivatives of the metric in it.
When backreaction is re-formulated as an effective fluid, these are
also the largest terms which appear~\cite{Baumann:2010tm}. In some
respects then this settles it: backreaction is small by virtue of
there being a very \emph{small} hierarchy of scales between the
Hubble scale at equality and the Hubble scale today (they are only a
factor of 50 apart in comoving terms). In this evaluation of
backreaction, then, what happens on scales smaller than the equality
scale is actually  of little relevance. This is perhaps surprising given how we
normally think of backreaction arising from small-scale structure~--
it is really power on very large scales which are responsible for
the backreaction effect.

On the other hand, however, we should be able describe the universe
using an orthonormal tetrad version of the Field Equations. This is
entirely equivalent to the EFE, but reformulates gravity as a system
of first-order PDE's in the Ricci Rotation coefficients and Weyl
curvature tensor. Because of this, metric variables occur at up to
two derivative levels higher than in the field equations. As we have
seen with $H$, taking its derivative in perturbation theory can
result in divergent terms appearing. Consequently, terms such as
$(\partial^2\Phi)^2$ will occur frequently, and as we have seen, in
$q$ at least, can appear in both their connected and non-connected
forms giving rise to large backreaction terms. That they can appear in purely 
kinematical quantities, and not just in a perhaps erroneous expansion of $\delta$, 
was not addressed in~\cite{Baumann:2010tm}. The non-connected
terms appear from the commutation relation for the time evolution of
the spatial domain, so may be an artefact of the non-covariant
averaging procedure.  The connected form, however, is much more
subtle to interpret as it appears even if we just calculate the
expectation value of $q$~-- there's nothing really to do with
averaging here. Where they appear in their connected form, they
depend on how we cut-off the non-linear scales (assuming
ergodicity). Rather intriguingly, a cutoff at the virial scale
yields significant changes to the background.  In this reading of
backreaction, then, backreaction could be very significant indeed, and is
large precisely because of the large hierarchy of scales between the
non-linear scale and the Hubble scale, and is dominated by the UV cutoff. The large equality scale
dampens the effect, but not enough for it to be insignificant.

Observationally, can there be any signature of backreaction? When
measuring the Hubble rate, perturbations are significant in the
variance of the Hubble rate on sub-equality
scales~\cite{Li:2007ci,Li:2008yj}. Perturbations affect the whole
distance-redshift relation which has been calculated to first-order
by~\cite{Bonvin:2005ps,Bonvin:2006en,Barausse:2005nf}. Corrections
to the luminosity distance include corrections $\partial^2\Phi$,
just like for the Hubble rate. This allows us to see that the
variance includes divergent terms like $(\partial^2\Phi)^2$~-- the
lensing term~-- which appear in the variance of the all-sky average
of the luminosity distance~-- see \cite{Bonvin:2005ps}, although
this problem was not discussed. 
It is an important open question to find out what happens if their
results are extended to second-order, where an overall ensemble
averaged offset to the luminosity distance will be present. Terms of
the form $(\partial^2\Phi)^2$ appear in the series expansion of the
distance-redshift relation, which can be used to define an
observational deceleration
parameter~\cite{Barausse:2005nf,Clarkson:2011uk}, but is this just
an artefact of using a power series?

Finally, Baumann et al~\cite{Baumann:2010tm} claim that there are significant effects on the BAO. This means that
backreaction effects are of significance to precision cosmology at
these scales, even if they are not significant on the largest
scales.

\section{The Alternatives}

We can observe why the averaging problem is so difficult in GR,
straight from the definition of the average of a
scalar~Eq.~(\ref{eq:Average}): to average or smooth a scalar
quantity $S$ it is not sufficient to just know $S(t,\bm x)$ as it
would be in other areas of physics which have a spacetime
prescribed; we also need to know $h_{ij}(t,\bm x)$ too which
requires the solution to the field equations in the nearby domain of
interest. To prescribe a distribution of matter with mean energy
density, say, requires us to know the full solution for the
spacetime~-- \emph{before} we can state its mean. Many of the
problems we have discussed stem from this simple fact: describing an
averaged spacetime accurately is just as difficult as modelling the
full lumpy spacetime. Combined with the fact that averaging tensors
covariantly is ill defined, the non-linearity of the field
equations and so on,   this is all quite a problem. That it is
difficult, everyone agrees; what its effects are, on the other hand,
depends on who is asking and what method they're using. Let us
summarise some of the different viewpoints.

\subsection{The Skeptic}

The view that backreaction is negligible has been strongly argued by
Peebles and Ratra~\cite{Peebles:2002gy,Peebles:2009hw}, Ishibashi
and Wald~\cite{Ishibashi:2005sj}, among quite a few
others~\cite{Flanagan:2005dk,Hirata:2005ei,Geshnizjani:2005ce,Bonvin:2005ps,
Bonvin:2006en,Behrend:2007mf,Kumar:2008uk,Baumann:2010tm,Green:2010qy}. The
backbone of their  argument is based on the fact that the
gravitational potential $\Phi$ is small everywhere in the universe
except in the immediate vicinity of  black holes, and so the average
of $\Phi$, and all \emph{relevant}  physical quantities derived from
it must consequently be small. While the density contrast can
fluctuate by many tens of orders-of-magnitude, it is not this that
causes backreaction because the field equations are linear in the
density. Relativistic backreaction is not caused by density
fluctuations \emph{per se}; rather, it arises due to peculiar
velocity contributions, which at the order relevant for backreaction
are $v^2 \sim (\partial\Phi)^2$, and are $\sim 10^{-5}$ in
dimensionless form. While density fluctuations behave as $\sim
\partial^2\Phi$ on small scales these can only cause a very large
\emph{variance} in cosmological parameters, and this is only on
scales which are small compared to the Hubble scale, about the
non-linear scale of a few Mpc.

Although there are non-linear interactions of the gravitational
potential with itself, this general relativistic effect of `gravity
gravitating' is tiny, $\sim \Phi^2\sim \mathcal{O}(10^{-10})$. So
backreaction and all perturbative effects on the background, while
interesting, are essentially irrelevant for cosmology.

\subsection{The Enthusiast}

Some argue that the view of the Skeptic misses the point in its
entirety: backreaction is a general relativistic effect arising from
the fully non-linear field equations \cite{Kolb:2005da,Barausse:2005nf,Wiltshire:2007jk,Wiltshire:2011CQG}. 
Arguments which claim that backreaction is small rely on perturbed
FLRW which are inherently quasi-Newtonian. Such models miss the main
possible backreaction effect both because they are Newtonian, and
because they don't take inhomogeneity seriously. Even though they
can be corrected to give some relativistic
effects~\cite{Chisari:2011iq}, this still only brings them into line
with linear relativistic perturbation theory, which does not
adequately capture the reality of a vast network of walls and filaments forming structures around expanding voids. 
 Furthermore,
Newtonian N-body simulations remove backreaction from the start
because they employ periodic boundary conditions enforced on the
background~\cite{Buchert:1995fz}. While such a condition is
enforced, there can't be a net flow of particles into or out of the
box~-- exactly what backreaction looks for~-- and the expansion is
simply put in by hand. The divergent terms from perturbation theory
$(\partial^2\Phi)^2$ cancel neatly in the backreaction terms because
of their Newtonian nature.

Because the averaged field equations do not easily close, the
backreaction terms have to be 
estimated from a realistic, fully non-linear, solution of the field
equations, which are hard to make realistic --- but that must be the
aim.

\subsection{The Fence-Sitter}

While backreaction is clearly an important mechanism to understand,
it seems difficult to believe it can really
cause significant effects on the largest cosmological scales, both
because of the large scale differences involved and the small value
of the gravitational potential. However, that does not in itself
imply that backreaction is insignificant in cosmology:
perturbations about FLRW do not necessarily give small effects on
smaller scales. Depending on the definition of the deceleration
parameter, for example, $(\partial^2\Phi)^2$ terms either appear or
cancel. They cancel in the deceleration parameter defined as the
deceleration of the averaged scale-factor; but in the average value
of the local deceleration parameter, they appear and give 10\%
effects even when power is artificially cutoff at 10~Mpc, and
diverge if power on smaller scales is allowed to
contribute~\cite{Clarkson:2009hr,Clarkson:2011uk}. This is hardly an unreasonable
thing to calculate, yet perturbation theory doesn't give a sensible
answer. Similar terms may appear at higher order in perturbation theory. Consequently, it seems too early to say that backreaction
and the associated fitting problem is negligible purely from
perturbative arguments.
Even a conservative calculation indicates it is
significant on BAO scales \cite{Baumann:2010tm}. That means it
cannot be ignored in precision cosmology.

\subsection{Open issues and future directions}

Perhaps we should ask ourselves the following questions, which may
be fruitful avenues for further research:

\begin{description}

\item[The role of $(\partial^2\Phi)^2$ at second-order] How should such terms be dealt with? Do they
signify that something unphysical has been calculated, or do they signify a breakdown in perturbation
theory? Do they appear in the luminosity distance at second-order and so give a large effect there? Or
is it that the ensemble average of such terms is actually irrelevant? Should perturbations be smoothed
on a given scale at each order before calculating the next, which can then go to smaller scales?

\item[Higher-order perturbation theory] Given the uncertainty in guessing what higher-order perturbation
theory might look like, it is timely to calculate directly, and to see if the perturbative expansion
remains small, and effectively truncates at second-order, or if it does indeed diverge and require
renormalisation methods.

\item[The validity of the coordinates]
How large a domain can be covered by the perturbative coordinates in
a spacetime with major voids~-- which are genuine vacuum and so static~-- combined with a crowd of
near-delta-functions for the density wherever there is any structure?
It is crucial to the usual arguments.

\item [Description of matter as dust] Cosmology doesn't deal
with many well defined particles. Smoothing over
    $10^{\sim70}$ CDM particles is well defined in kinetic theory to give a fluid
    description, but once a structure freezes out of the expansion, it becomes our
    `point particle', in an ill defined sense. Then we have $10^{\sim12}$ virialised
    structures~-- significantly less than Avagadro's number~-- which are large
    compared to the Hubble volume, and large compared to the total mass, compared to normal fluids. Are these correctly modelled as dust? Within perturbation theory it appears consistent~\cite{Baumann:2010tm}; can this be made more general~\cite{Wiltshire:2011vy}?
    How does this relate to the radiation-reaction problem~\cite{Poisson:2003nc}?

\item[Relativistic simulations]
Is a fully general relativistic simulation out of the question? This would have to be attempted
very differently from current N-body simulations, but could be attempted on the expectation of
improving computing power. Rather than work with point particles, one could start with the ADM
equations with different fluids, plus perturbations laid down early. A simulation could consist
of perturbative modes up to a maximum wavenumber, rather than N particles of a given mass. The
resolution would be determined by the maximum wavenumber, so simulations would start big and get
smaller as computing power improved.

\item[Averaging observables] The role of observables and the lightcone is still very unclear in
the averaging scenario. Observable quantities such as distances are
usually implicitly sky-averaged quantities. While it is not clear
what an averaged lightcone might mean, sky-averages of observables
such as the redshift-distance relation have a reasonably clear
meaning, even in a very non-homogeneous universe. How does this fit
into other approaches? 

\item [Covariant formulation] The present detailed analyses are highly gauge dependent.
Initial work on gauge invariant formalism is interesting
\cite{Gasperini:2009mu,AbrBraMuk97} but ideally one would use a 1+3
covariant and gauge invariant geometric formalism for the purpose. Alternatively, a tetrad approach may be useful as  all the Ricci rotation coefficients, Weyl tensors and so on are then scalar
invariants. How does one average the tetrad?

\end{description}



\section{Conclusion}

Averaging, even though it implicitly defines the standard
concordance cosmology, is not understood in general relativity.
Because extra terms in the field equations seem inevitable, then
it is surely timely to understand how
to deal with it~-- or if we can't, then decide what limitations that
imposes on cosmology.

We have seen in this article some of the large variety of ways that
this problem is being investigated. Some argue it's utterly
irrelevant, suppressed by orders of magnitude below the background
model which consequently holds genuine physical significance~--
analogous to being able to say ``The Earth is approximately a
sphere". Others take the opposite view and contend that it holds the
key to the whole dark energy issue. More moderate views prevail
between these two extremes, pointing out there may be significant
effects in intermediate scales of cosmological interest. It is also
fair to point to studies where simplified but fully non-linear
spacetimes have observational properties significantly different
from their spatially averaged counterparts \cite{Rasanen:2004js,Paranjape:2006cd}. And it is not clear that
perturbation theory is genuinely convergent, and that it is a well
behaved process. Is it properly converged at second order, where
claims for tiny backreaction are made? Why are we so lucky that the
equality scale is so large, suppressing what could be a huge
backreaction effect? Why, in short, is cosmology so easy?

At the very least, then, these considerations surely tell us that it
is important to understand the averaging, fitting and backreaction
problems 
to see what  effects there may be on cosmology.
There \emph{are} some scales where backreaction may be important
- probably not the largest scales relevant to the cosmic
acceleration, but others where precision cosmology is significant.
In investigating this, we must get a clearer distinction between
dynamical and observational effects - the latter not covered here,
but certainly relevant to null fitting, which is the core of
observational cosmology.


\acknowledgments

 We thank Thomas Buchert, Alan Coley, Ruth
Durrer, Syksy R\"as\"anen, and David Wiltshire for comments and
discussions, and an anonymous reviewer for useful comments that have
helped shape the text. This work is funded in part by the NRF (South
Africa). OU is funded by the University of Cape Town, the National
Institute of Theoretical Physics (NITheP, South Africa) and the
Square Kilometre Array (SKA, South Africa).

\newpage
\bibliography{cosmoref}

\begin{thebibliography}{136}
\expandafter\ifx\csname natexlab\endcsname\relax\def\natexlab#1{#1}\fi
\expandafter\ifx\csname bibnamefont\endcsname\relax
  \def\bibnamefont#1{#1}\fi
\expandafter\ifx\csname bibfnamefont\endcsname\relax
  \def\bibfnamefont#1{#1}\fi
\expandafter\ifx\csname citenamefont\endcsname\relax
  \def\citenamefont#1{#1}\fi
\expandafter\ifx\csname url\endcsname\relax
  \def\url#1{\texttt{#1}}\fi
\expandafter\ifx\csname urlprefix\endcsname\relax\def\urlprefix{URL }\fi
\providecommand{\bibinfo}[2]{#2}
\providecommand{\eprint}[2][]{\url{#2}}

\bibitem[{\citenamefont{Buchert}(2000)}]{Buchert:1999er}
\bibinfo{author}{\bibfnamefont{T.}~\bibnamefont{Buchert}},
  \bibinfo{journal}{Gen. Rel. Grav.} \textbf{\bibinfo{volume}{32}},
  \bibinfo{pages}{105} (\bibinfo{year}{2000}), \eprint{gr-qc/9906015}.

\bibitem[{\citenamefont{Buchert}(2001)}]{Buchert:2001sa}
\bibinfo{author}{\bibfnamefont{T.}~\bibnamefont{Buchert}},
  \bibinfo{journal}{Gen. Rel. Grav.} \textbf{\bibinfo{volume}{33}},
  \bibinfo{pages}{1381} (\bibinfo{year}{2001}), \eprint{gr-qc/0102049}.

\bibitem[{\citenamefont{Rasanen}(2004{\natexlab{a}})}]{Rasanen:2003fy}
\bibinfo{author}{\bibfnamefont{S.}~\bibnamefont{Rasanen}},
  \bibinfo{journal}{JCAP} \textbf{\bibinfo{volume}{0402}}, \bibinfo{pages}{003}
  (\bibinfo{year}{2004}{\natexlab{a}}), \eprint{astro-ph/0311257}.

\bibitem[{\citenamefont{Barausse et~al.}(2005)\citenamefont{Barausse,
  Matarrese, and Riotto}}]{Barausse:2005nf}
\bibinfo{author}{\bibfnamefont{E.}~\bibnamefont{Barausse}},
  \bibinfo{author}{\bibfnamefont{S.}~\bibnamefont{Matarrese}},
  \bibnamefont{and} \bibinfo{author}{\bibfnamefont{A.}~\bibnamefont{Riotto}},
  \bibinfo{journal}{Phys. Rev.} \textbf{\bibinfo{volume}{D71}},
  \bibinfo{pages}{063537} (\bibinfo{year}{2005}), \eprint{astro-ph/0501152}.

\bibitem[{\citenamefont{Kolb et~al.}(2006)\citenamefont{Kolb, Matarrese, and
  Riotto}}]{Kolb:2005da}
\bibinfo{author}{\bibfnamefont{E.~W.} \bibnamefont{Kolb}},
  \bibinfo{author}{\bibfnamefont{S.}~\bibnamefont{Matarrese}},
  \bibnamefont{and} \bibinfo{author}{\bibfnamefont{A.}~\bibnamefont{Riotto}},
  \bibinfo{journal}{New J. Phys.} \textbf{\bibinfo{volume}{8}},
  \bibinfo{pages}{322} (\bibinfo{year}{2006}), \eprint{astro-ph/0506534}.

\bibitem[{\citenamefont{Rasanen}(2006)}]{Rasanen:2006kp}
\bibinfo{author}{\bibfnamefont{S.}~\bibnamefont{Rasanen}},
  \bibinfo{journal}{JCAP} \textbf{\bibinfo{volume}{0611}}, \bibinfo{pages}{003}
  (\bibinfo{year}{2006}), \eprint{astro-ph/0607626}.

\bibitem[{\citenamefont{Kasai}(2007)}]{Kasai:2007fn}
\bibinfo{author}{\bibfnamefont{M.}~\bibnamefont{Kasai}},
  \bibinfo{journal}{Prog. Theor. Phys.} \textbf{\bibinfo{volume}{117}},
  \bibinfo{pages}{1067} (\bibinfo{year}{2007}), \eprint{astro-ph/0703298}.

\bibitem[{\citenamefont{Ishibashi and Wald}(2006)}]{Ishibashi:2005sj}
\bibinfo{author}{\bibfnamefont{A.}~\bibnamefont{Ishibashi}} \bibnamefont{and}
  \bibinfo{author}{\bibfnamefont{R.~M.} \bibnamefont{Wald}},
  \bibinfo{journal}{Class. Quant. Grav.} \textbf{\bibinfo{volume}{23}},
  \bibinfo{pages}{235} (\bibinfo{year}{2006}), \eprint{gr-qc/0509108}.

\bibitem[{\citenamefont{Flanagan}(2005)}]{Flanagan:2005dk}
\bibinfo{author}{\bibfnamefont{E.~E.} \bibnamefont{Flanagan}},
  \bibinfo{journal}{Phys. Rev.} \textbf{\bibinfo{volume}{D71}},
  \bibinfo{pages}{103521} (\bibinfo{year}{2005}), \eprint{hep-th/0503202}.

\bibitem[{\citenamefont{Hirata and Seljak}(2005)}]{Hirata:2005ei}
\bibinfo{author}{\bibfnamefont{C.~M.} \bibnamefont{Hirata}} \bibnamefont{and}
  \bibinfo{author}{\bibfnamefont{U.}~\bibnamefont{Seljak}},
  \bibinfo{journal}{Phys. Rev.} \textbf{\bibinfo{volume}{D72}},
  \bibinfo{pages}{083501} (\bibinfo{year}{2005}), \eprint{astro-ph/0503582}.

\bibitem[{\citenamefont{Geshnizjani et~al.}(2005)\citenamefont{Geshnizjani,
  Chung, and Afshordi}}]{Geshnizjani:2005ce}
\bibinfo{author}{\bibfnamefont{G.}~\bibnamefont{Geshnizjani}},
  \bibinfo{author}{\bibfnamefont{D.~J.~H.} \bibnamefont{Chung}},
  \bibnamefont{and} \bibinfo{author}{\bibfnamefont{N.}~\bibnamefont{Afshordi}},
  \bibinfo{journal}{Phys. Rev.} \textbf{\bibinfo{volume}{D72}},
  \bibinfo{pages}{023517} (\bibinfo{year}{2005}), \eprint{astro-ph/0503553}.

\bibitem[{\citenamefont{Bonvin et~al.}(2006{\natexlab{a}})\citenamefont{Bonvin,
  Durrer, and Gasperini}}]{Bonvin:2005ps}
\bibinfo{author}{\bibfnamefont{C.}~\bibnamefont{Bonvin}},
  \bibinfo{author}{\bibfnamefont{R.}~\bibnamefont{Durrer}}, \bibnamefont{and}
  \bibinfo{author}{\bibfnamefont{M.}~\bibnamefont{Gasperini}},
  \bibinfo{journal}{Phys.Rev.} \textbf{\bibinfo{volume}{D73}},
  \bibinfo{pages}{023523} (\bibinfo{year}{2006}{\natexlab{a}}),
  \eprint{astro-ph/0511183}.

\bibitem[{\citenamefont{Bonvin et~al.}(2006{\natexlab{b}})\citenamefont{Bonvin,
  Durrer, and Kunz}}]{Bonvin:2006en}
\bibinfo{author}{\bibfnamefont{C.}~\bibnamefont{Bonvin}},
  \bibinfo{author}{\bibfnamefont{R.}~\bibnamefont{Durrer}}, \bibnamefont{and}
  \bibinfo{author}{\bibfnamefont{M.}~\bibnamefont{Kunz}},
  \bibinfo{journal}{Phys.Rev.Lett.} \textbf{\bibinfo{volume}{96}},
  \bibinfo{pages}{191302} (\bibinfo{year}{2006}{\natexlab{b}}),
  \eprint{astro-ph/0603240}.

\bibitem[{\citenamefont{Behrend et~al.}(2008)\citenamefont{Behrend, Brown, and
  Robbers}}]{Behrend:2007mf}
\bibinfo{author}{\bibfnamefont{J.}~\bibnamefont{Behrend}},
  \bibinfo{author}{\bibfnamefont{I.~A.} \bibnamefont{Brown}}, \bibnamefont{and}
  \bibinfo{author}{\bibfnamefont{G.}~\bibnamefont{Robbers}},
  \bibinfo{journal}{JCAP} \textbf{\bibinfo{volume}{0801}}, \bibinfo{pages}{013}
  (\bibinfo{year}{2008}), \eprint{0710.4964}.

\bibitem[{\citenamefont{Kumar and Flanagan}(2008)}]{Kumar:2008uk}
\bibinfo{author}{\bibfnamefont{N.}~\bibnamefont{Kumar}} \bibnamefont{and}
  \bibinfo{author}{\bibfnamefont{E.~E.} \bibnamefont{Flanagan}},
  \bibinfo{journal}{Phys. Rev.} \textbf{\bibinfo{volume}{D78}},
  \bibinfo{pages}{063537} (\bibinfo{year}{2008}), \eprint{0808.1043}.

\bibitem[{\citenamefont{Krasinski et~al.}(2010)\citenamefont{Krasinski,
  Hellaby, Celerier, and Bolejko}}]{Krasinski:2009qq}
\bibinfo{author}{\bibfnamefont{A.}~\bibnamefont{Krasinski}},
  \bibinfo{author}{\bibfnamefont{C.}~\bibnamefont{Hellaby}},
  \bibinfo{author}{\bibfnamefont{M.-N.} \bibnamefont{Celerier}},
  \bibnamefont{and} \bibinfo{author}{\bibfnamefont{K.}~\bibnamefont{Bolejko}},
  \bibinfo{journal}{Gen.Rel.Grav.} \textbf{\bibinfo{volume}{42}},
  \bibinfo{pages}{2453} (\bibinfo{year}{2010}), \eprint{0903.4070}.

\bibitem[{\citenamefont{{Tomita}}(2009)}]{Tomita:2009ar}
\bibinfo{author}{\bibfnamefont{K.}~\bibnamefont{{Tomita}}},
  \bibinfo{journal}{ArXiv e-prints}  (\bibinfo{year}{2009}),
  \eprint{0906.1325}.

\bibitem[{\citenamefont{{Baumann} et~al.}(2010)\citenamefont{{Baumann},
  {Nicolis}, {Senatore}, and {Zaldarriaga}}}]{Baumann:2010tm}
\bibinfo{author}{\bibfnamefont{D.}~\bibnamefont{{Baumann}}},
  \bibinfo{author}{\bibfnamefont{A.}~\bibnamefont{{Nicolis}}},
  \bibinfo{author}{\bibfnamefont{L.}~\bibnamefont{{Senatore}}},
  \bibnamefont{and}
  \bibinfo{author}{\bibfnamefont{M.}~\bibnamefont{{Zaldarriaga}}},
  \bibinfo{journal}{ArXiv e-prints}  (\bibinfo{year}{2010}),
  \eprint{1004.2488}.

\bibitem[{\citenamefont{{Green} and {Wald}}(2010)}]{Green:2010qy}
\bibinfo{author}{\bibfnamefont{S.~R.} \bibnamefont{{Green}}} \bibnamefont{and}
  \bibinfo{author}{\bibfnamefont{R.~M.} \bibnamefont{{Wald}}},
  \bibinfo{journal}{ArXiv e-prints}  (\bibinfo{year}{2010}),
  \eprint{1011.4920}.

\bibitem[{\citenamefont{Russ et~al.}(1997)\citenamefont{Russ, Soffel, Kasai,
  and Borner}}]{Russ:1996km}
\bibinfo{author}{\bibfnamefont{H.}~\bibnamefont{Russ}},
  \bibinfo{author}{\bibfnamefont{M.~H.} \bibnamefont{Soffel}},
  \bibinfo{author}{\bibfnamefont{M.}~\bibnamefont{Kasai}}, \bibnamefont{and}
  \bibinfo{author}{\bibfnamefont{G.}~\bibnamefont{Borner}},
  \bibinfo{journal}{Phys. Rev.} \textbf{\bibinfo{volume}{D56}},
  \bibinfo{pages}{2044} (\bibinfo{year}{1997}), \eprint{astro-ph/9612218}.

\bibitem[{\citenamefont{Kolb et~al.}(2005)\citenamefont{Kolb, Matarrese,
  Notari, and Riotto}}]{Kolb:2004am}
\bibinfo{author}{\bibfnamefont{E.~W.} \bibnamefont{Kolb}},
  \bibinfo{author}{\bibfnamefont{S.}~\bibnamefont{Matarrese}},
  \bibinfo{author}{\bibfnamefont{A.}~\bibnamefont{Notari}}, \bibnamefont{and}
  \bibinfo{author}{\bibfnamefont{A.}~\bibnamefont{Riotto}},
  \bibinfo{journal}{Phys. Rev.} \textbf{\bibinfo{volume}{D71}},
  \bibinfo{pages}{023524} (\bibinfo{year}{2005}), \eprint{hep-ph/0409038}.

\bibitem[{\citenamefont{Li and Schwarz}(2007)}]{Li:2007ci}
\bibinfo{author}{\bibfnamefont{N.}~\bibnamefont{Li}} \bibnamefont{and}
  \bibinfo{author}{\bibfnamefont{D.~J.} \bibnamefont{Schwarz}},
  \bibinfo{journal}{Phys. Rev.} \textbf{\bibinfo{volume}{D76}},
  \bibinfo{pages}{083011} (\bibinfo{year}{2007}), \eprint{gr-qc/0702043}.

\bibitem[{\citenamefont{Li and Schwarz}(2008)}]{Li:2007ny}
\bibinfo{author}{\bibfnamefont{N.}~\bibnamefont{Li}} \bibnamefont{and}
  \bibinfo{author}{\bibfnamefont{D.~J.} \bibnamefont{Schwarz}},
  \bibinfo{journal}{Phys. Rev.} \textbf{\bibinfo{volume}{D78}},
  \bibinfo{pages}{083531} (\bibinfo{year}{2008}), \eprint{0710.5073}.

\bibitem[{\citenamefont{Li et~al.}(2008)\citenamefont{Li, Seikel, and
  Schwarz}}]{Li:2008yj}
\bibinfo{author}{\bibfnamefont{N.}~\bibnamefont{Li}},
  \bibinfo{author}{\bibfnamefont{M.}~\bibnamefont{Seikel}}, \bibnamefont{and}
  \bibinfo{author}{\bibfnamefont{D.~J.} \bibnamefont{Schwarz}},
  \bibinfo{journal}{Fortsch. Phys.} \textbf{\bibinfo{volume}{56}},
  \bibinfo{pages}{465} (\bibinfo{year}{2008}), \eprint{0801.3420}.

\bibitem[{\citenamefont{Clarkson et~al.}(2009)\citenamefont{Clarkson, Ananda,
  and Larena}}]{Clarkson:2009hr}
\bibinfo{author}{\bibfnamefont{C.}~\bibnamefont{Clarkson}},
  \bibinfo{author}{\bibfnamefont{K.}~\bibnamefont{Ananda}}, \bibnamefont{and}
  \bibinfo{author}{\bibfnamefont{J.}~\bibnamefont{Larena}},
  \bibinfo{journal}{Phys. Rev.} \textbf{\bibinfo{volume}{D80}},
  \bibinfo{pages}{083525} (\bibinfo{year}{2009}), \eprint{0907.3377}.

\bibitem[{\citenamefont{Clarkson}(2010)}]{Clarkson:2009jq}
\bibinfo{author}{\bibfnamefont{C.}~\bibnamefont{Clarkson}},
  \bibinfo{journal}{AIP Conf.Proc.} \textbf{\bibinfo{volume}{1241}},
  \bibinfo{pages}{784} (\bibinfo{year}{2010}), \eprint{0911.2601}.

\bibitem[{\citenamefont{{Chung}}(2010)}]{Chung:2010xx}
\bibinfo{author}{\bibfnamefont{H.}~\bibnamefont{{Chung}}},
  \bibinfo{journal}{ArXiv e-prints}  (\bibinfo{year}{2010}),
  \eprint{1009.1333}.

\bibitem[{\citenamefont{{Umeh} et~al.}(2010)\citenamefont{{Umeh}, {Larena}, and
  {Clarkson}}}]{Umeh:2010pr}
\bibinfo{author}{\bibfnamefont{O.}~\bibnamefont{{Umeh}}},
  \bibinfo{author}{\bibfnamefont{J.}~\bibnamefont{{Larena}}}, \bibnamefont{and}
  \bibinfo{author}{\bibfnamefont{C.}~\bibnamefont{{Clarkson}}},
  \bibinfo{journal}{ArXiv e-prints}  (\bibinfo{year}{2010}),
  \eprint{1011.3959}.

\bibitem[{\citenamefont{Wang et~al.}(1998)\citenamefont{Wang, Spergel, and
  Turner}}]{Wang:1997tp}
\bibinfo{author}{\bibfnamefont{Y.}~\bibnamefont{Wang}},
  \bibinfo{author}{\bibfnamefont{D.~N.} \bibnamefont{Spergel}},
  \bibnamefont{and} \bibinfo{author}{\bibfnamefont{E.~L.}
  \bibnamefont{Turner}}, \bibinfo{journal}{Astrophys. J.}
  \textbf{\bibinfo{volume}{498}}, \bibinfo{pages}{1} (\bibinfo{year}{1998}),
  \eprint{astro-ph/9708014}.

\bibitem[{\citenamefont{Shi and Turner}(1998)}]{Shi:1997aa}
\bibinfo{author}{\bibfnamefont{X.-D.} \bibnamefont{Shi}} \bibnamefont{and}
  \bibinfo{author}{\bibfnamefont{M.~S.} \bibnamefont{Turner}},
  \bibinfo{journal}{Astrophys. J.} \textbf{\bibinfo{volume}{493}},
  \bibinfo{pages}{519} (\bibinfo{year}{1998}), \eprint{astro-ph/9707101}.

\bibitem[{\citenamefont{{Bolejko}}(2011)}]{Bolejko:2011ej}
\bibinfo{author}{\bibfnamefont{K.}~\bibnamefont{{Bolejko}}},
  \bibinfo{journal}{ArXiv e-prints}  (\bibinfo{year}{2011}),
  \eprint{1101.3338}.

\bibitem[{\citenamefont{Kolb}(to appear)}]{Kolb:2011}
\bibinfo{author}{\bibfnamefont{E.~W.} \bibnamefont{Kolb}},
  \bibinfo{journal}{Classical and Quantum Gravity}  (\bibinfo{year}{to
  appear}).

\bibitem[{\citenamefont{{Kolb} et~al.}(2005)\citenamefont{{Kolb}, {Matarrese},
  {Notari}, and {Riotto}}}]{Kolb:2005me}
\bibinfo{author}{\bibfnamefont{E.~W.} \bibnamefont{{Kolb}}},
  \bibinfo{author}{\bibfnamefont{S.}~\bibnamefont{{Matarrese}}},
  \bibinfo{author}{\bibfnamefont{A.}~\bibnamefont{{Notari}}}, \bibnamefont{and}
  \bibinfo{author}{\bibfnamefont{A.}~\bibnamefont{{Riotto}}},
  \bibinfo{journal}{ArXiv e-prints}  (\bibinfo{year}{2005}),
  \eprint{arXiv:hep-th/0503117}.

\bibitem[{\citenamefont{Wiltshire}(2009{\natexlab{a}})}]{Wiltshire:2009ip}
\bibinfo{author}{\bibfnamefont{D.~L.} \bibnamefont{Wiltshire}},
  \bibinfo{journal}{EAS Publ. Ser.} \textbf{\bibinfo{volume}{36}},
  \bibinfo{pages}{91} (\bibinfo{year}{2009}{\natexlab{a}}), \eprint{0912.5234}.

\bibitem[{\citenamefont{Wiltshire}(2007{\natexlab{a}})}]{Wiltshire:2007jk}
\bibinfo{author}{\bibfnamefont{D.~L.} \bibnamefont{Wiltshire}},
  \bibinfo{journal}{New J. Phys.} \textbf{\bibinfo{volume}{9}},
  \bibinfo{pages}{377} (\bibinfo{year}{2007}{\natexlab{a}}),
  \eprint{gr-qc/0702082}.

\bibitem[{\citenamefont{{Abramo} et~al.}(1997)\citenamefont{{Abramo},
  {Brandenberger}, and {Mukhanov}}}]{AbrBraMuk97}
\bibinfo{author}{\bibfnamefont{L.~R.~W.} \bibnamefont{{Abramo}}},
  \bibinfo{author}{\bibfnamefont{R.~H.} \bibnamefont{{Brandenberger}}},
  \bibnamefont{and} \bibinfo{author}{\bibfnamefont{V.~F.}
  \bibnamefont{{Mukhanov}}}, \bibinfo{journal}{\prd}
  \textbf{\bibinfo{volume}{56}}, \bibinfo{pages}{3248} (\bibinfo{year}{1997}),
  \eprint{arXiv:gr-qc/9704037}.

\bibitem[{\citenamefont{{Calzetta} et~al.}(2001)\citenamefont{{Calzetta}, {Hu},
  and {Mazzitelli}}}]{CalHuMaz01}
\bibinfo{author}{\bibfnamefont{E.~A.} \bibnamefont{{Calzetta}}},
  \bibinfo{author}{\bibfnamefont{B.~L.} \bibnamefont{{Hu}}}, \bibnamefont{and}
  \bibinfo{author}{\bibfnamefont{F.~D.} \bibnamefont{{Mazzitelli}}},
  \bibinfo{journal}{\physrep} \textbf{\bibinfo{volume}{352}},
  \bibinfo{pages}{459} (\bibinfo{year}{2001}), \eprint{arXiv:hep-th/0102199}.

\bibitem[{\citenamefont{Labini and Pietronero}(2010)}]{Labini:2010aj}
\bibinfo{author}{\bibfnamefont{F.~S.} \bibnamefont{Labini}} \bibnamefont{and}
  \bibinfo{author}{\bibfnamefont{L.}~\bibnamefont{Pietronero}},
  \bibinfo{journal}{J. Stat. Mech.} \textbf{\bibinfo{volume}{2010}},
  \bibinfo{pages}{11029} (\bibinfo{year}{2010}), \eprint{1012.5624}.

\bibitem[{\citenamefont{Labini}(2010)}]{Labini:2010mg}
\bibinfo{author}{\bibfnamefont{F.~S.} \bibnamefont{Labini}},
  \bibinfo{journal}{Astron. Astrophys.} \textbf{\bibinfo{volume}{523}},
  \bibinfo{pages}{A68} (\bibinfo{year}{2010}), \eprint{1007.1860}.

\bibitem[{\citenamefont{{Springel} et~al.}(2005)\citenamefont{{Springel},
  {White}, {Jenkins}, {Frenk}, {Yoshida}, {Gao}, {Navarro}, {Thacker},
  {Croton}, {Helly} et~al.}}]{MilSim2005}
\bibinfo{author}{\bibfnamefont{V.}~\bibnamefont{{Springel}}},
  \bibinfo{author}{\bibfnamefont{S.~D.~M.} \bibnamefont{{White}}},
  \bibinfo{author}{\bibfnamefont{A.}~\bibnamefont{{Jenkins}}},
  \bibinfo{author}{\bibfnamefont{C.~S.} \bibnamefont{{Frenk}}},
  \bibinfo{author}{\bibfnamefont{N.}~\bibnamefont{{Yoshida}}},
  \bibinfo{author}{\bibfnamefont{L.}~\bibnamefont{{Gao}}},
  \bibinfo{author}{\bibfnamefont{J.}~\bibnamefont{{Navarro}}},
  \bibinfo{author}{\bibfnamefont{R.}~\bibnamefont{{Thacker}}},
  \bibinfo{author}{\bibfnamefont{D.}~\bibnamefont{{Croton}}},
  \bibinfo{author}{\bibfnamefont{J.}~\bibnamefont{{Helly}}},
  \bibnamefont{et~al.}, \bibinfo{journal}{\nat} \textbf{\bibinfo{volume}{435}},
  \bibinfo{pages}{629} (\bibinfo{year}{2005}), \eprint{arXiv:astro-ph/0504097}.

\bibitem[{\citenamefont{{Ellis} and {Stoeger}}(1987)}]{EllSto87}
\bibinfo{author}{\bibfnamefont{G.~F.~R.} \bibnamefont{{Ellis}}}
  \bibnamefont{and}
  \bibinfo{author}{\bibfnamefont{W.}~\bibnamefont{{Stoeger}}},
  \bibinfo{journal}{Class. Quant. Grav.} \textbf{\bibinfo{volume}{4}},
  \bibinfo{pages}{1697} (\bibinfo{year}{1987}).

\bibitem[{\citenamefont{Isaacson}(1968)}]{Isa68a}
\bibinfo{author}{\bibfnamefont{R.~A.} \bibnamefont{Isaacson}},
  \bibinfo{journal}{Phys. Rev.} \textbf{\bibinfo{volume}{166}},
  \bibinfo{pages}{1272} (\bibinfo{year}{1968}).

\bibitem[{\citenamefont{{Isaacson}}(1968)}]{Isa68b}
\bibinfo{author}{\bibfnamefont{R.~A.} \bibnamefont{{Isaacson}}},
  \bibinfo{journal}{Physical Review} \textbf{\bibinfo{volume}{166}},
  \bibinfo{pages}{1263} (\bibinfo{year}{1968}).

\bibitem[{\citenamefont{{MacCallum} and {Taub}}(1973)}]{MacTau73}
\bibinfo{author}{\bibfnamefont{M.~A.~H.} \bibnamefont{{MacCallum}}}
  \bibnamefont{and} \bibinfo{author}{\bibfnamefont{A.~H.}
  \bibnamefont{{Taub}}}, \bibinfo{journal}{Communications in Mathematical
  Physics} \textbf{\bibinfo{volume}{30}}, \bibinfo{pages}{153}
  (\bibinfo{year}{1973}).

\bibitem[{\citenamefont{{Ellis}}(1984)}]{Ell84}
\bibinfo{author}{\bibfnamefont{G.~F.~R.} \bibnamefont{{Ellis}}}, in
  \emph{\bibinfo{booktitle}{General Relativity and Gravitation Conference}},
  edited by \bibinfo{editor}{\bibnamefont{{B.~Bertotti, F.~de Felice, \&
  A.~Pascolini}}} (\bibinfo{year}{1984}), pp. \bibinfo{pages}{215--288}.

\bibitem[{\citenamefont{{Kristian} and {Sachs}}(1966)}]{KandS66}
\bibinfo{author}{\bibfnamefont{J.}~\bibnamefont{{Kristian}}} \bibnamefont{and}
  \bibinfo{author}{\bibfnamefont{R.~K.} \bibnamefont{{Sachs}}},
  \bibinfo{journal}{\apj} \textbf{\bibinfo{volume}{143}}, \bibinfo{pages}{379}
  (\bibinfo{year}{1966}).

\bibitem[{\citenamefont{{Ellis} and {MacCallum}}(1969)}]{EllisMac69}
\bibinfo{author}{\bibfnamefont{G.~F.~R.} \bibnamefont{{Ellis}}}
  \bibnamefont{and} \bibinfo{author}{\bibfnamefont{M.~A.~H.}
  \bibnamefont{{MacCallum}}}, \bibinfo{journal}{Communications in Mathematical
  Physics} \textbf{\bibinfo{volume}{12}}, \bibinfo{pages}{108}
  (\bibinfo{year}{1969}).

\bibitem[{\citenamefont{{MacCallum} and {Ellis}}(1970)}]{MacEllis70}
\bibinfo{author}{\bibfnamefont{M.~A.~H.} \bibnamefont{{MacCallum}}}
  \bibnamefont{and} \bibinfo{author}{\bibfnamefont{G.~F.~R.}
  \bibnamefont{{Ellis}}}, \bibinfo{journal}{Communications in Mathematical
  Physics} \textbf{\bibinfo{volume}{19}}, \bibinfo{pages}{31}
  (\bibinfo{year}{1970}).

\bibitem[{\citenamefont{{Ellis} et~al.}(1985)\citenamefont{{Ellis}, {Nel},
  {Maartens}, {Stoeger}, and {Whitman}}}]{Elletal85}
\bibinfo{author}{\bibfnamefont{G.~F.~R.} \bibnamefont{{Ellis}}},
  \bibinfo{author}{\bibfnamefont{S.~D.} \bibnamefont{{Nel}}},
  \bibinfo{author}{\bibfnamefont{R.}~\bibnamefont{{Maartens}}},
  \bibinfo{author}{\bibfnamefont{W.~R.} \bibnamefont{{Stoeger}}},
  \bibnamefont{and} \bibinfo{author}{\bibfnamefont{A.~P.}
  \bibnamefont{{Whitman}}}, \bibinfo{journal}{\physrep}
  \textbf{\bibinfo{volume}{124}}, \bibinfo{pages}{315} (\bibinfo{year}{1985}).

\bibitem[{\citenamefont{Clarkson}(2000)}]{Clarksonthesis}
\bibinfo{author}{\bibfnamefont{C.~A.} \bibnamefont{Clarkson}},
  \bibinfo{journal}{ArXiv Astrophysics e-prints}  (\bibinfo{year}{2000}),
  \eprint{arXiv:astro-ph/0008089}.

\bibitem[{\citenamefont{Bertortti}(1966)}]{Bertotii:1966}
\bibinfo{author}{\bibfnamefont{B.}~\bibnamefont{Bertortti}},
  \bibinfo{journal}{Proc. Roy. Soc. London} \textbf{\bibinfo{volume}{A, 294}},
  \bibinfo{pages}{195} (\bibinfo{year}{1966}), \eprint{astro-ph}.

\bibitem[{\citenamefont{{Dyer} and {Roeder}}(1974)}]{Dyer-Roeder}
\bibinfo{author}{\bibfnamefont{C.~C.} \bibnamefont{{Dyer}}} \bibnamefont{and}
  \bibinfo{author}{\bibfnamefont{R.~C.} \bibnamefont{{Roeder}}},
  \bibinfo{journal}{\apj} \textbf{\bibinfo{volume}{189}}, \bibinfo{pages}{167}
  (\bibinfo{year}{1974}).

\bibitem[{\citenamefont{{Coley}}(2009)}]{Coley:2009qc}
\bibinfo{author}{\bibfnamefont{A.~A.} \bibnamefont{{Coley}}},
  \bibinfo{journal}{ArXiv e-prints}  (\bibinfo{year}{2009}),
  \eprint{0905.2442}.

\bibitem[{\citenamefont{Rasanen}(2010{\natexlab{a}})}]{Rasanen:2009uw}
\bibinfo{author}{\bibfnamefont{S.}~\bibnamefont{Rasanen}},
  \bibinfo{journal}{JCAP} \textbf{\bibinfo{volume}{1003}}, \bibinfo{pages}{018}
  (\bibinfo{year}{2010}{\natexlab{a}}), \eprint{0912.3370}.

\bibitem[{\citenamefont{Rasanen}(2009)}]{Rasanen:2008be}
\bibinfo{author}{\bibfnamefont{S.}~\bibnamefont{Rasanen}},
  \bibinfo{journal}{JCAP} \textbf{\bibinfo{volume}{0902}}, \bibinfo{pages}{011}
  (\bibinfo{year}{2009}), \eprint{0812.2872}.

\bibitem[{\citenamefont{{Gasperini} et~al.}(2011)\citenamefont{{Gasperini},
  {Marozzi}, {Nugier}, and {Veneziano}}}]{Gasperini:2011us}
\bibinfo{author}{\bibfnamefont{M.}~\bibnamefont{{Gasperini}}},
  \bibinfo{author}{\bibfnamefont{G.}~\bibnamefont{{Marozzi}}},
  \bibinfo{author}{\bibfnamefont{F.}~\bibnamefont{{Nugier}}}, \bibnamefont{and}
  \bibinfo{author}{\bibfnamefont{G.}~\bibnamefont{{Veneziano}}},
  \bibinfo{journal}{ArXiv e-prints}  (\bibinfo{year}{2011}),
  \eprint{1104.1167}.

\bibitem[{\citenamefont{{Ellis} and {Stoeger}}(2009)}]{EllSto09}
\bibinfo{author}{\bibfnamefont{G.~F.~R.} \bibnamefont{{Ellis}}}
  \bibnamefont{and} \bibinfo{author}{\bibfnamefont{W.~R.}
  \bibnamefont{{Stoeger}}}, \bibinfo{journal}{\mnras}
  \textbf{\bibinfo{volume}{398}}, \bibinfo{pages}{1527} (\bibinfo{year}{2009}),
  \eprint{1001.4572}.

\bibitem[{\citenamefont{Buchert}(2008)}]{Buchert:2007ik}
\bibinfo{author}{\bibfnamefont{T.}~\bibnamefont{Buchert}},
  \bibinfo{journal}{Gen. Rel. Grav.} \textbf{\bibinfo{volume}{40}},
  \bibinfo{pages}{467} (\bibinfo{year}{2008}), \eprint{0707.2153}.

\bibitem[{\citenamefont{Zalaletdinov}(1997)}]{Zalaletdinov:1996aj}
\bibinfo{author}{\bibfnamefont{R.~M.} \bibnamefont{Zalaletdinov}},
  \bibinfo{journal}{Bull. Astron. Soc. India} \textbf{\bibinfo{volume}{25}},
  \bibinfo{pages}{401} (\bibinfo{year}{1997}), \eprint{gr-qc/9703016}.

\bibitem[{\citenamefont{Zalaletdinov}(2008)}]{Zalaletdinov:2008ts}
\bibinfo{author}{\bibfnamefont{R.}~\bibnamefont{Zalaletdinov}},
  \bibinfo{journal}{Int. J. Mod. Phys.} \textbf{\bibinfo{volume}{A23}},
  \bibinfo{pages}{1173} (\bibinfo{year}{2008}), \eprint{0801.3256}.

\bibitem[{\citenamefont{Wiltshire}(2007{\natexlab{b}})}]{Wiltshire:2007fg}
\bibinfo{author}{\bibfnamefont{D.~L.} \bibnamefont{Wiltshire}},
  \bibinfo{journal}{Phys. Rev. Lett.} \textbf{\bibinfo{volume}{99}},
  \bibinfo{pages}{251101} (\bibinfo{year}{2007}{\natexlab{b}}),
  \eprint{0709.0732}.

\bibitem[{\citenamefont{{Zotov} and
  {Stoeger}}(1992{\natexlab{a}})}]{1992CQGra...9.1023Z}
\bibinfo{author}{\bibfnamefont{N.~V.} \bibnamefont{{Zotov}}} \bibnamefont{and}
  \bibinfo{author}{\bibfnamefont{W.~R.} \bibnamefont{{Stoeger}}},
  \bibinfo{journal}{Classical and Quantum Gravity}
  \textbf{\bibinfo{volume}{9}}, \bibinfo{pages}{1023}
  (\bibinfo{year}{1992}{\natexlab{a}}).

\bibitem[{\citenamefont{{Zotov} and
  {Stoeger}}(1992{\natexlab{b}})}]{1992AAS...180.5804Z}
\bibinfo{author}{\bibfnamefont{N.~V.} \bibnamefont{{Zotov}}} \bibnamefont{and}
  \bibinfo{author}{\bibfnamefont{W.~R.} \bibnamefont{{Stoeger}}}, in
  \emph{\bibinfo{booktitle}{American Astronomical Society Meeting Abstracts
  \#180}} (\bibinfo{year}{1992}{\natexlab{b}}), vol.~\bibinfo{volume}{24} of
  \emph{\bibinfo{series}{Bulletin of the American Astronomical Society}}, pp.
  \bibinfo{pages}{822--+}.

\bibitem[{\citenamefont{{Zotov} and {Stoeger}}(1995)}]{1995ApJ...453..574Z}
\bibinfo{author}{\bibfnamefont{N.~V.} \bibnamefont{{Zotov}}} \bibnamefont{and}
  \bibinfo{author}{\bibfnamefont{W.~R.} \bibnamefont{{Stoeger}}},
  \bibinfo{journal}{\apj} \textbf{\bibinfo{volume}{453}}, \bibinfo{pages}{574}
  (\bibinfo{year}{1995}).

\bibitem[{\citenamefont{{van den Hoogen}}(2010)}]{Hoogen:2010qb}
\bibinfo{author}{\bibfnamefont{R.~J.} \bibnamefont{{van den Hoogen}}},
  \bibinfo{journal}{ArXiv e-prints}  (\bibinfo{year}{2010}),
  \eprint{1003.4020}.

\bibitem[{\citenamefont{Brannlund et~al.}(2010)\citenamefont{Brannlund, Hoogen,
  and Coley}}]{Brannlund:2010rs}
\bibinfo{author}{\bibfnamefont{J.}~\bibnamefont{Brannlund}},
  \bibinfo{author}{\bibfnamefont{R.~v.~d.} \bibnamefont{Hoogen}},
  \bibnamefont{and} \bibinfo{author}{\bibfnamefont{A.}~\bibnamefont{Coley}},
  \bibinfo{journal}{Int.J.Mod.Phys.} \textbf{\bibinfo{volume}{D19}},
  \bibinfo{pages}{1915} (\bibinfo{year}{2010}), \eprint{1003.2014}.

\bibitem[{\citenamefont{Carfora and Piotrkowska}(1995)}]{Carfora:1995fj}
\bibinfo{author}{\bibfnamefont{M.}~\bibnamefont{Carfora}} \bibnamefont{and}
  \bibinfo{author}{\bibfnamefont{K.}~\bibnamefont{Piotrkowska}},
  \bibinfo{journal}{Phys. Rev.} \textbf{\bibinfo{volume}{D52}},
  \bibinfo{pages}{4393} (\bibinfo{year}{1995}), \eprint{gr-qc/9502021}.

\bibitem[{\citenamefont{Futamase}(1996)}]{Futamase:1996fk}
\bibinfo{author}{\bibfnamefont{T.}~\bibnamefont{Futamase}},
  \bibinfo{journal}{Phys.Rev.} \textbf{\bibinfo{volume}{D53}},
  \bibinfo{pages}{681} (\bibinfo{year}{1996}).

\bibitem[{\citenamefont{Boersma}(1998)}]{Boersma:1997yt}
\bibinfo{author}{\bibfnamefont{J.~P.} \bibnamefont{Boersma}},
  \bibinfo{journal}{Phys.Rev.} \textbf{\bibinfo{volume}{D57}},
  \bibinfo{pages}{798} (\bibinfo{year}{1998}), \eprint{gr-qc/9711057}.

\bibitem[{\citenamefont{Stoeger et~al.}(2007)\citenamefont{Stoeger, Helmi, and
  Torres}}]{Stoeger:1999ig}
\bibinfo{author}{\bibfnamefont{S.}~\bibnamefont{Stoeger},
  \bibfnamefont{William~R.}},
  \bibinfo{author}{\bibfnamefont{A.}~\bibnamefont{Helmi}}, \bibnamefont{and}
  \bibinfo{author}{\bibfnamefont{D.~F.} \bibnamefont{Torres}},
  \bibinfo{journal}{Int.J.Mod.Phys.} \textbf{\bibinfo{volume}{D16}},
  \bibinfo{pages}{1001} (\bibinfo{year}{2007}), \eprint{gr-qc/9904020}.

\bibitem[{\citenamefont{Buchert and Ehlers}(1997)}]{Buchert:1995fz}
\bibinfo{author}{\bibfnamefont{T.}~\bibnamefont{Buchert}} \bibnamefont{and}
  \bibinfo{author}{\bibfnamefont{J.}~\bibnamefont{Ehlers}},
  \bibinfo{journal}{Astron.Astrophys.} \textbf{\bibinfo{volume}{320}},
  \bibinfo{pages}{1} (\bibinfo{year}{1997}), \eprint{astro-ph/9510056}.

\bibitem[{\citenamefont{{Buchert}}(1996)}]{Buchert:1995pj}
\bibinfo{author}{\bibfnamefont{T.}~\bibnamefont{{Buchert}}}, in
  \emph{\bibinfo{booktitle}{Mapping, Measuring, and Modelling the Universe}},
  edited by \bibinfo{editor}{\bibnamefont{{P.~Coles, V.~Martinez, \&
  M.-J.~Pons-Borderia}}} (\bibinfo{year}{1996}), vol.~\bibinfo{volume}{94} of
  \emph{\bibinfo{series}{Astronomical Society of the Pacific Conference
  Series}}, pp. \bibinfo{pages}{349--+}, \eprint{arXiv:astro-ph/9512107}.

\bibitem[{\citenamefont{Larena}(2009)}]{Larena:2009md}
\bibinfo{author}{\bibfnamefont{J.}~\bibnamefont{Larena}},
  \bibinfo{journal}{Phys. Rev.} \textbf{\bibinfo{volume}{D79}},
  \bibinfo{pages}{084006} (\bibinfo{year}{2009}), \eprint{0902.3159}.

\bibitem[{\citenamefont{Gasperini et~al.}(2010)\citenamefont{Gasperini,
  Marozzi, and Veneziano}}]{Gasperini:2009mu}
\bibinfo{author}{\bibfnamefont{M.}~\bibnamefont{Gasperini}},
  \bibinfo{author}{\bibfnamefont{G.}~\bibnamefont{Marozzi}}, \bibnamefont{and}
  \bibinfo{author}{\bibfnamefont{G.}~\bibnamefont{Veneziano}},
  \bibinfo{journal}{JCAP} \textbf{\bibinfo{volume}{1002}}, \bibinfo{pages}{009}
  (\bibinfo{year}{2010}), \eprint{0912.3244}.

\bibitem[{\citenamefont{{Nambu} and {Tanimoto}}(2005)}]{Nambu:2005zn}
\bibinfo{author}{\bibfnamefont{Y.}~\bibnamefont{{Nambu}}} \bibnamefont{and}
  \bibinfo{author}{\bibfnamefont{M.}~\bibnamefont{{Tanimoto}}},
  \bibinfo{journal}{ArXiv General Relativity and Quantum Cosmology e-prints}
  (\bibinfo{year}{2005}), \eprint{arXiv:gr-qc/0507057}.

\bibitem[{\citenamefont{Moffat}(2006)}]{Moffat:2005ii}
\bibinfo{author}{\bibfnamefont{J.~W.} \bibnamefont{Moffat}},
  \bibinfo{journal}{JCAP} \textbf{\bibinfo{volume}{0605}}, \bibinfo{pages}{001}
  (\bibinfo{year}{2006}), \eprint{astro-ph/0505326}.

\bibitem[{\citenamefont{Wiegand and Buchert}(2010)}]{Wiegand:2010uh}
\bibinfo{author}{\bibfnamefont{A.}~\bibnamefont{Wiegand}} \bibnamefont{and}
  \bibinfo{author}{\bibfnamefont{T.}~\bibnamefont{Buchert}},
  \bibinfo{journal}{Phys. Rev.} \textbf{\bibinfo{volume}{D82}},
  \bibinfo{pages}{023523} (\bibinfo{year}{2010}), \eprint{1002.3912}.

\bibitem[{\citenamefont{Coley}(2010)}]{Coley:2009yz}
\bibinfo{author}{\bibfnamefont{A.}~\bibnamefont{Coley}},
  \bibinfo{journal}{Class.Quant.Grav.} \textbf{\bibinfo{volume}{27}},
  \bibinfo{pages}{245017} (\bibinfo{year}{2010}), \eprint{0908.4281}.

\bibitem[{\citenamefont{Zalaletdinov}(1993)}]{Zalaletdinov:1992cf}
\bibinfo{author}{\bibfnamefont{R.}~\bibnamefont{Zalaletdinov}},
  \bibinfo{journal}{Gen.Rel.Grav.} \textbf{\bibinfo{volume}{25}},
  \bibinfo{pages}{673} (\bibinfo{year}{1993}).

\bibitem[{\citenamefont{Zalaletdinov}(1992)}]{Zalaletdinov:1992cg}
\bibinfo{author}{\bibfnamefont{R.~M.} \bibnamefont{Zalaletdinov}},
  \bibinfo{journal}{Gen.Rel.Grav.} \textbf{\bibinfo{volume}{24}},
  \bibinfo{pages}{1015} (\bibinfo{year}{1992}).

\bibitem[{\citenamefont{Zalaletdinov}(2004)}]{Zalaletdinov:2004wd}
\bibinfo{author}{\bibfnamefont{R.}~\bibnamefont{Zalaletdinov}},
  \bibinfo{journal}{Annals Eur.Acad.Sci.}  (\bibinfo{year}{2004}),
  \eprint{gr-qc/0411004}.

\bibitem[{\citenamefont{Paranjape and
  Singh}(2008{\natexlab{a}})}]{Paranjape:2006ww}
\bibinfo{author}{\bibfnamefont{A.}~\bibnamefont{Paranjape}} \bibnamefont{and}
  \bibinfo{author}{\bibfnamefont{T.~P.} \bibnamefont{Singh}},
  \bibinfo{journal}{Gen. Rel. Grav.} \textbf{\bibinfo{volume}{40}},
  \bibinfo{pages}{139} (\bibinfo{year}{2008}{\natexlab{a}}),
  \eprint{astro-ph/0609481}.

\bibitem[{\citenamefont{{Montani} and {Zalaletdinov}}(1999)}]{Montani:1999}
\bibinfo{author}{\bibfnamefont{G.}~\bibnamefont{{Montani}}} \bibnamefont{and}
  \bibinfo{author}{\bibfnamefont{R.}~\bibnamefont{{Zalaletdinov}}}, in
  \emph{\bibinfo{booktitle}{Recent Developments in Theoretical and Experimental
  General Relativity, Gravitation, and Relativistic Field Theories}}, edited by
  \bibinfo{editor}{\bibnamefont{{T.~Piran \& R.~Ruffini}}}
  (\bibinfo{year}{1999}), pp. \bibinfo{pages}{628--+}.

\bibitem[{\citenamefont{Coley et~al.}(2005)\citenamefont{Coley, Pelavas, and
  Zalaletdinov}}]{Coley:2005ei}
\bibinfo{author}{\bibfnamefont{A.~A.} \bibnamefont{Coley}},
  \bibinfo{author}{\bibfnamefont{N.}~\bibnamefont{Pelavas}}, \bibnamefont{and}
  \bibinfo{author}{\bibfnamefont{R.~M.} \bibnamefont{Zalaletdinov}},
  \bibinfo{journal}{Phys. Rev. Lett.} \textbf{\bibinfo{volume}{95}},
  \bibinfo{pages}{151102} (\bibinfo{year}{2005}), \eprint{gr-qc/0504115}.

\bibitem[{\citenamefont{Coley and Pelavas}(2006)}]{Coley:2006xu}
\bibinfo{author}{\bibfnamefont{A.}~\bibnamefont{Coley}} \bibnamefont{and}
  \bibinfo{author}{\bibfnamefont{N.}~\bibnamefont{Pelavas}},
  \bibinfo{journal}{Phys.Rev.} \textbf{\bibinfo{volume}{D74}},
  \bibinfo{pages}{087301} (\bibinfo{year}{2006}), \eprint{astro-ph/0606535}.

\bibitem[{\citenamefont{Coley and Pelavas}(2007)}]{Coley:2006kp}
\bibinfo{author}{\bibfnamefont{A.}~\bibnamefont{Coley}} \bibnamefont{and}
  \bibinfo{author}{\bibfnamefont{N.}~\bibnamefont{Pelavas}},
  \bibinfo{journal}{Phys.Rev.} \textbf{\bibinfo{volume}{D75}},
  \bibinfo{pages}{043506} (\bibinfo{year}{2007}), \eprint{gr-qc/0607079}.

\bibitem[{\citenamefont{van~den Hoogen}(2009)}]{vandenHoogen:2009nh}
\bibinfo{author}{\bibfnamefont{R.}~\bibnamefont{van~den Hoogen}},
  \bibinfo{journal}{J.Math.Phys.} \textbf{\bibinfo{volume}{50}},
  \bibinfo{pages}{082503} (\bibinfo{year}{2009}), \eprint{0909.0070}.

\bibitem[{\citenamefont{Paranjape and
  Singh}(2007{\natexlab{a}})}]{Paranjape:2007wr}
\bibinfo{author}{\bibfnamefont{A.}~\bibnamefont{Paranjape}} \bibnamefont{and}
  \bibinfo{author}{\bibfnamefont{T.}~\bibnamefont{Singh}},
  \bibinfo{journal}{Phys.Rev.} \textbf{\bibinfo{volume}{D76}},
  \bibinfo{pages}{044006} (\bibinfo{year}{2007}{\natexlab{a}}),
  \eprint{gr-qc/0703106}.

\bibitem[{\citenamefont{Paranjape}(2008)}]{Paranjape:2008mx}
\bibinfo{author}{\bibfnamefont{A.}~\bibnamefont{Paranjape}},
  \bibinfo{journal}{Phys. Rev.} \textbf{\bibinfo{volume}{D78}},
  \bibinfo{pages}{063522} (\bibinfo{year}{2008}), \eprint{0806.2755}.

\bibitem[{\citenamefont{Paranjape and
  Singh}(2008{\natexlab{b}})}]{Paranjape:2008ai}
\bibinfo{author}{\bibfnamefont{A.}~\bibnamefont{Paranjape}} \bibnamefont{and}
  \bibinfo{author}{\bibfnamefont{T.~P.} \bibnamefont{Singh}},
  \bibinfo{journal}{JCAP} \textbf{\bibinfo{volume}{0803}}, \bibinfo{pages}{023}
  (\bibinfo{year}{2008}{\natexlab{b}}), \eprint{0801.1546}.

\bibitem[{\citenamefont{Buchert and Carfora}(2002)}]{Buchert:2002ht}
\bibinfo{author}{\bibfnamefont{T.}~\bibnamefont{Buchert}} \bibnamefont{and}
  \bibinfo{author}{\bibfnamefont{M.}~\bibnamefont{Carfora}},
  \bibinfo{journal}{Class. Quant. Grav.} \textbf{\bibinfo{volume}{19}},
  \bibinfo{pages}{6109} (\bibinfo{year}{2002}), \eprint{gr-qc/0210037}.

\bibitem[{\citenamefont{Buchert and Carfora}(2003)}]{Buchert:2002ij}
\bibinfo{author}{\bibfnamefont{T.}~\bibnamefont{Buchert}} \bibnamefont{and}
  \bibinfo{author}{\bibfnamefont{M.}~\bibnamefont{Carfora}},
  \bibinfo{journal}{Phys. Rev. Lett.} \textbf{\bibinfo{volume}{90}},
  \bibinfo{pages}{031101} (\bibinfo{year}{2003}), \eprint{gr-qc/0210045}.

\bibitem[{\citenamefont{Wiltshire}(2008)}]{Wiltshire:2008sg}
\bibinfo{author}{\bibfnamefont{D.~L.} \bibnamefont{Wiltshire}},
  \bibinfo{journal}{Phys. Rev.} \textbf{\bibinfo{volume}{D78}},
  \bibinfo{pages}{084032} (\bibinfo{year}{2008}), \eprint{0809.1183}.

\bibitem[{\citenamefont{Wiltshire}(2009{\natexlab{b}})}]{Wiltshire:2009db}
\bibinfo{author}{\bibfnamefont{D.~L.} \bibnamefont{Wiltshire}},
  \bibinfo{journal}{Phys. Rev.} \textbf{\bibinfo{volume}{D80}},
  \bibinfo{pages}{123512} (\bibinfo{year}{2009}{\natexlab{b}}),
  \eprint{0909.0749}.

\bibitem[{\citenamefont{Leith et~al.}(2008)\citenamefont{Leith, Ng, and
  Wiltshire}}]{Leith:2007ay}
\bibinfo{author}{\bibfnamefont{B.~M.} \bibnamefont{Leith}},
  \bibinfo{author}{\bibfnamefont{S.~C.~C.} \bibnamefont{Ng}}, \bibnamefont{and}
  \bibinfo{author}{\bibfnamefont{D.~L.} \bibnamefont{Wiltshire}},
  \bibinfo{journal}{Astrophys. J.} \textbf{\bibinfo{volume}{672}},
  \bibinfo{pages}{L91} (\bibinfo{year}{2008}), \eprint{0709.2535}.

\bibitem[{\citenamefont{Mattsson and Mattsson}(2010)}]{Mattsson:2010vq}
\bibinfo{author}{\bibfnamefont{M.}~\bibnamefont{Mattsson}} \bibnamefont{and}
  \bibinfo{author}{\bibfnamefont{T.}~\bibnamefont{Mattsson}},
  \bibinfo{journal}{JCAP} \textbf{\bibinfo{volume}{1010}}, \bibinfo{pages}{021}
  (\bibinfo{year}{2010}), \eprint{1007.2939}.

\bibitem[{\citenamefont{{Mattsson} and {Mattsson}}(2010)}]{Mattsson:2010ky}
\bibinfo{author}{\bibfnamefont{M.}~\bibnamefont{{Mattsson}}} \bibnamefont{and}
  \bibinfo{author}{\bibfnamefont{T.}~\bibnamefont{{Mattsson}}},
  \bibinfo{journal}{ArXiv e-prints}  (\bibinfo{year}{2010}),
  \eprint{1012.4008}.

\bibitem[{\citenamefont{{Einstein} and {Straus}}(1945)}]{EinsteinandStraus}
\bibinfo{author}{\bibfnamefont{A.}~\bibnamefont{{Einstein}}} \bibnamefont{and}
  \bibinfo{author}{\bibfnamefont{E.~G.} \bibnamefont{{Straus}}},
  \bibinfo{journal}{Reviews of Modern Physics} \textbf{\bibinfo{volume}{17}},
  \bibinfo{pages}{120} (\bibinfo{year}{1945}).

\bibitem[{\citenamefont{{Kantowski}}(1969)}]{Kantowski1969}
\bibinfo{author}{\bibfnamefont{R.}~\bibnamefont{{Kantowski}}},
  \bibinfo{journal}{\apj} \textbf{\bibinfo{volume}{155}}, \bibinfo{pages}{89}
  (\bibinfo{year}{1969}).

\bibitem[{\citenamefont{Tomita}(2000)}]{Tomita:1999qn}
\bibinfo{author}{\bibfnamefont{K.}~\bibnamefont{Tomita}},
  \bibinfo{journal}{Astrophys.J.} \textbf{\bibinfo{volume}{529}},
  \bibinfo{pages}{38} (\bibinfo{year}{2000}), \eprint{astro-ph/9906027}.

\bibitem[{\citenamefont{Hellaby and Krasinski}(2006)}]{Hellaby:2005ut}
\bibinfo{author}{\bibfnamefont{C.}~\bibnamefont{Hellaby}} \bibnamefont{and}
  \bibinfo{author}{\bibfnamefont{A.}~\bibnamefont{Krasinski}},
  \bibinfo{journal}{Phys.Rev.} \textbf{\bibinfo{volume}{D73}},
  \bibinfo{pages}{023518} (\bibinfo{year}{2006}), \eprint{gr-qc/0510093}.

\bibitem[{\citenamefont{Biswas and Notari}(2008)}]{Biswas:2007gi}
\bibinfo{author}{\bibfnamefont{T.}~\bibnamefont{Biswas}} \bibnamefont{and}
  \bibinfo{author}{\bibfnamefont{A.}~\bibnamefont{Notari}},
  \bibinfo{journal}{JCAP} \textbf{\bibinfo{volume}{0806}}, \bibinfo{pages}{021}
  (\bibinfo{year}{2008}), \eprint{astro-ph/0702555}.

\bibitem[{\citenamefont{Vanderveld et~al.}(2006)\citenamefont{Vanderveld,
  Flanagan, and Wasserman}}]{Vanderveld:2006rb}
\bibinfo{author}{\bibfnamefont{R.}~\bibnamefont{Vanderveld}},
  \bibinfo{author}{\bibfnamefont{E.~E.} \bibnamefont{Flanagan}},
  \bibnamefont{and}
  \bibinfo{author}{\bibfnamefont{I.}~\bibnamefont{Wasserman}},
  \bibinfo{journal}{Phys.Rev.} \textbf{\bibinfo{volume}{D74}},
  \bibinfo{pages}{023506} (\bibinfo{year}{2006}), \eprint{astro-ph/0602476}.

\bibitem[{\citenamefont{Marra et~al.}(2007)\citenamefont{Marra, Kolb,
  Matarrese, and Riotto}}]{Marra:2007pm}
\bibinfo{author}{\bibfnamefont{V.}~\bibnamefont{Marra}},
  \bibinfo{author}{\bibfnamefont{E.~W.} \bibnamefont{Kolb}},
  \bibinfo{author}{\bibfnamefont{S.}~\bibnamefont{Matarrese}},
  \bibnamefont{and} \bibinfo{author}{\bibfnamefont{A.}~\bibnamefont{Riotto}},
  \bibinfo{journal}{Phys.Rev.} \textbf{\bibinfo{volume}{D76}},
  \bibinfo{pages}{123004} (\bibinfo{year}{2007}), \eprint{0708.3622}.

\bibitem[{\citenamefont{Marra et~al.}(2008)\citenamefont{Marra, Kolb, and
  Matarrese}}]{Marra:2007gc}
\bibinfo{author}{\bibfnamefont{V.}~\bibnamefont{Marra}},
  \bibinfo{author}{\bibfnamefont{E.~W.} \bibnamefont{Kolb}}, \bibnamefont{and}
  \bibinfo{author}{\bibfnamefont{S.}~\bibnamefont{Matarrese}},
  \bibinfo{journal}{Phys.Rev.} \textbf{\bibinfo{volume}{D77}},
  \bibinfo{pages}{023003} (\bibinfo{year}{2008}), \eprint{0710.5505}.

\bibitem[{\citenamefont{Kolb et~al.}(2010)\citenamefont{Kolb, Marra, and
  Matarrese}}]{Kolb:2009rp}
\bibinfo{author}{\bibfnamefont{E.~W.} \bibnamefont{Kolb}},
  \bibinfo{author}{\bibfnamefont{V.}~\bibnamefont{Marra}}, \bibnamefont{and}
  \bibinfo{author}{\bibfnamefont{S.}~\bibnamefont{Matarrese}},
  \bibinfo{journal}{Gen.Rel.Grav.} \textbf{\bibinfo{volume}{42}},
  \bibinfo{pages}{1399} (\bibinfo{year}{2010}), \eprint{0901.4566}.

\bibitem[{\citenamefont{Vanderveld et~al.}(2008)\citenamefont{Vanderveld,
  Flanagan, and Wasserman}}]{Vanderveld:2008vi}
\bibinfo{author}{\bibfnamefont{R.}~\bibnamefont{Vanderveld}},
  \bibinfo{author}{\bibfnamefont{E.~E.} \bibnamefont{Flanagan}},
  \bibnamefont{and}
  \bibinfo{author}{\bibfnamefont{I.}~\bibnamefont{Wasserman}},
  \bibinfo{journal}{Phys.Rev.} \textbf{\bibinfo{volume}{D78}},
  \bibinfo{pages}{083511} (\bibinfo{year}{2008}), \eprint{0808.1080}.

\bibitem[{\citenamefont{Sugiura et~al.}(2000)\citenamefont{Sugiura, Nakao, Ida,
  Sakai, and Ishihara}}]{Sugiura:1999fm}
\bibinfo{author}{\bibfnamefont{N.}~\bibnamefont{Sugiura}},
  \bibinfo{author}{\bibfnamefont{K.-i.} \bibnamefont{Nakao}},
  \bibinfo{author}{\bibfnamefont{D.}~\bibnamefont{Ida}},
  \bibinfo{author}{\bibfnamefont{N.}~\bibnamefont{Sakai}}, \bibnamefont{and}
  \bibinfo{author}{\bibfnamefont{H.}~\bibnamefont{Ishihara}},
  \bibinfo{journal}{Prog. Theor. Phys.} \textbf{\bibinfo{volume}{103}},
  \bibinfo{pages}{73} (\bibinfo{year}{2000}), \eprint{astro-ph/9912414}.

\bibitem[{\citenamefont{{Clifton} and {Zuntz}}(2009)}]{Clifton:2009nv}
\bibinfo{author}{\bibfnamefont{T.}~\bibnamefont{{Clifton}}} \bibnamefont{and}
  \bibinfo{author}{\bibfnamefont{J.}~\bibnamefont{{Zuntz}}},
  \bibinfo{journal}{\mnras} \textbf{\bibinfo{volume}{400}},
  \bibinfo{pages}{2185} (\bibinfo{year}{2009}), \eprint{0902.0726}.

\bibitem[{\citenamefont{Bolejko and Celerier}(2010)}]{Bolejko:2010eb}
\bibinfo{author}{\bibfnamefont{K.}~\bibnamefont{Bolejko}} \bibnamefont{and}
  \bibinfo{author}{\bibfnamefont{M.-N.} \bibnamefont{Celerier}},
  \bibinfo{journal}{Phys.Rev.} \textbf{\bibinfo{volume}{D82}},
  \bibinfo{pages}{103510} (\bibinfo{year}{2010}), \eprint{1005.2584}.

\bibitem[{\citenamefont{{Szybka}}(2010)}]{Szybka:2010ky}
\bibinfo{author}{\bibfnamefont{S.~J.} \bibnamefont{{Szybka}}},
  \bibinfo{journal}{ArXiv e-prints}  (\bibinfo{year}{2010}),
  \eprint{1012.5239}.

\bibitem[{\citenamefont{Bolejko}(2009)}]{Bolejko:2008xh}
\bibinfo{author}{\bibfnamefont{K.}~\bibnamefont{Bolejko}},
  \bibinfo{journal}{Gen.Rel.Grav.} \textbf{\bibinfo{volume}{41}},
  \bibinfo{pages}{1737} (\bibinfo{year}{2009}), \eprint{0804.1846}.

\bibitem[{\citenamefont{Linquist and Wheeler}(1957)}]{LindWheel}
\bibinfo{author}{\bibfnamefont{R.~W.} \bibnamefont{Linquist}} \bibnamefont{and}
  \bibinfo{author}{\bibfnamefont{J.~A.} \bibnamefont{Wheeler}},
  \bibinfo{journal}{Rev. Mod. Phys} \textbf{\bibinfo{volume}{29}},
  \bibinfo{pages}{432} (\bibinfo{year}{1957}).

\bibitem[{\citenamefont{{Clifton}}(2010)}]{Clifton:2010fr}
\bibinfo{author}{\bibfnamefont{T.}~\bibnamefont{{Clifton}}},
  \bibinfo{journal}{ArXiv e-prints}  (\bibinfo{year}{2010}),
  \eprint{1005.0788}.

\bibitem[{\citenamefont{Clifton and Ferreira}(2009)}]{Clifton:2009jw}
\bibinfo{author}{\bibfnamefont{T.}~\bibnamefont{Clifton}} \bibnamefont{and}
  \bibinfo{author}{\bibfnamefont{P.~G.} \bibnamefont{Ferreira}},
  \bibinfo{journal}{Phys. Rev.} \textbf{\bibinfo{volume}{D80}},
  \bibinfo{pages}{103503} (\bibinfo{year}{2009}), \eprint{0907.4109}.

\bibitem[{\citenamefont{Paranjape}(2009)}]{Paranjape:2009zu}
\bibinfo{author}{\bibfnamefont{A.}~\bibnamefont{Paranjape}}
  (\bibinfo{year}{2009}), \bibinfo{note}{* Brief entry *}, \eprint{0906.3165}.

\bibitem[{\citenamefont{Rasanen}(2010{\natexlab{b}})}]{Rasanen:2010wz}
\bibinfo{author}{\bibfnamefont{S.}~\bibnamefont{Rasanen}},
  \bibinfo{journal}{Phys. Rev.} \textbf{\bibinfo{volume}{D81}},
  \bibinfo{pages}{103512} (\bibinfo{year}{2010}{\natexlab{b}}),
  \eprint{1002.4779}.

\bibitem[{\citenamefont{{Clarkson} and {Umeh}}(2011)}]{Clarkson:2011uk}
\bibinfo{author}{\bibfnamefont{C.}~\bibnamefont{{Clarkson}}} \bibnamefont{and}
  \bibinfo{author}{\bibfnamefont{O.}~\bibnamefont{{Umeh}}},
  \bibinfo{journal}{ArXiv e-prints}  (\bibinfo{year}{2011}),
  \eprint{1105.1886}.

\bibitem[{\citenamefont{{Noonan}}(1984)}]{Noonan1984}
\bibinfo{author}{\bibfnamefont{T.~W.} \bibnamefont{{Noonan}}},
  \bibinfo{journal}{General Relativity and Gravitation}
  \textbf{\bibinfo{volume}{16}}, \bibinfo{pages}{1103} (\bibinfo{year}{1984}).

\bibitem[{\citenamefont{{Noonan}}(1985)}]{Noonan1985}
\bibinfo{author}{\bibfnamefont{T.~W.} \bibnamefont{{Noonan}}},
  \bibinfo{journal}{General Relativity and Gravitation}
  \textbf{\bibinfo{volume}{17}}, \bibinfo{pages}{535} (\bibinfo{year}{1985}).

\bibitem[{\citenamefont{Paranjape and
  Singh}(2007{\natexlab{b}})}]{Paranjape:2007uj}
\bibinfo{author}{\bibfnamefont{A.}~\bibnamefont{Paranjape}} \bibnamefont{and}
  \bibinfo{author}{\bibfnamefont{T.}~\bibnamefont{Singh}},
  \bibinfo{journal}{Phys.Rev.} \textbf{\bibinfo{volume}{D76}},
  \bibinfo{pages}{044006} (\bibinfo{year}{2007}{\natexlab{b}}),
  \eprint{gr-qc/0703106}.

\bibitem[{\citenamefont{Brown et~al.}(2009)\citenamefont{Brown, Robbers, and
  Behrend}}]{Brown:2008ra}
\bibinfo{author}{\bibfnamefont{I.~A.} \bibnamefont{Brown}},
  \bibinfo{author}{\bibfnamefont{G.}~\bibnamefont{Robbers}}, \bibnamefont{and}
  \bibinfo{author}{\bibfnamefont{J.}~\bibnamefont{Behrend}},
  \bibinfo{journal}{JCAP} \textbf{\bibinfo{volume}{0904}}, \bibinfo{pages}{016}
  (\bibinfo{year}{2009}), \eprint{0811.4495}.

\bibitem[{\citenamefont{Peebles}(2010)}]{Peebles:2009hw}
\bibinfo{author}{\bibfnamefont{P.}~\bibnamefont{Peebles}},
  \bibinfo{journal}{AIP Conf.Proc.} \textbf{\bibinfo{volume}{1241}},
  \bibinfo{pages}{175} (\bibinfo{year}{2010}), \eprint{0910.5142}.

\bibitem[{\citenamefont{Bartolo et~al.}(2006)\citenamefont{Bartolo, Matarrese,
  and Riotto}}]{Bartolo:2005kv}
\bibinfo{author}{\bibfnamefont{N.}~\bibnamefont{Bartolo}},
  \bibinfo{author}{\bibfnamefont{S.}~\bibnamefont{Matarrese}},
  \bibnamefont{and} \bibinfo{author}{\bibfnamefont{A.}~\bibnamefont{Riotto}},
  \bibinfo{journal}{JCAP} \textbf{\bibinfo{volume}{0605}}, \bibinfo{pages}{010}
  (\bibinfo{year}{2006}), \eprint{astro-ph/0512481}.

\bibitem[{\citenamefont{Noh et~al.}(2009)\citenamefont{Noh, Jeong, and
  Hwang}}]{Noh:2009yu}
\bibinfo{author}{\bibfnamefont{H.}~\bibnamefont{Noh}},
  \bibinfo{author}{\bibfnamefont{D.}~\bibnamefont{Jeong}}, \bibnamefont{and}
  \bibinfo{author}{\bibfnamefont{J.-c.} \bibnamefont{Hwang}},
  \bibinfo{journal}{Phys. Rev. Lett.} \textbf{\bibinfo{volume}{103}},
  \bibinfo{pages}{021301} (\bibinfo{year}{2009}), \eprint{0902.4285}.

\bibitem[{\citenamefont{Notari}(2006)}]{Notari:2005xk}
\bibinfo{author}{\bibfnamefont{A.}~\bibnamefont{Notari}},
  \bibinfo{journal}{Mod. Phys. Lett.} \textbf{\bibinfo{volume}{A21}},
  \bibinfo{pages}{2997} (\bibinfo{year}{2006}), \eprint{astro-ph/0503715}.

\bibitem[{\citenamefont{Crocce and Scoccimarro}(2006)}]{Crocce:2005xy}
\bibinfo{author}{\bibfnamefont{M.}~\bibnamefont{Crocce}} \bibnamefont{and}
  \bibinfo{author}{\bibfnamefont{R.}~\bibnamefont{Scoccimarro}},
  \bibinfo{journal}{Phys.Rev.} \textbf{\bibinfo{volume}{D73}},
  \bibinfo{pages}{063519} (\bibinfo{year}{2006}), \eprint{astro-ph/0509418}.

\bibitem[{\citenamefont{{Gourgoulhon} and
  {Bonazzola}}(1994)}]{Gourgoulhon:1994}
\bibinfo{author}{\bibfnamefont{E.}~\bibnamefont{{Gourgoulhon}}}
  \bibnamefont{and}
  \bibinfo{author}{\bibfnamefont{S.}~\bibnamefont{{Bonazzola}}},
  \bibinfo{journal}{Classical and Quantum Gravity}
  \textbf{\bibinfo{volume}{11}}, \bibinfo{pages}{443} (\bibinfo{year}{1994}).

\bibitem[{\citenamefont{{Burnett}}(1989)}]{Burnette:1988}
\bibinfo{author}{\bibfnamefont{G.~A.} \bibnamefont{{Burnett}}},
  \bibinfo{journal}{Journal of Mathematical Physics}
  \textbf{\bibinfo{volume}{30}}, \bibinfo{pages}{90} (\bibinfo{year}{1989}).

\bibitem[{\citenamefont{Peebles and Ratra}(2003)}]{Peebles:2002gy}
\bibinfo{author}{\bibfnamefont{P.~J.~E.} \bibnamefont{Peebles}}
  \bibnamefont{and} \bibinfo{author}{\bibfnamefont{B.}~\bibnamefont{Ratra}},
  \bibinfo{journal}{Rev. Mod. Phys.} \textbf{\bibinfo{volume}{75}},
  \bibinfo{pages}{559} (\bibinfo{year}{2003}), \eprint{astro-ph/0207347}.

\bibitem[{\citenamefont{{Wiltshire}}(2011)}]{Wiltshire:2011CQG}
\bibinfo{author}{\bibfnamefont{D.~L.} \bibnamefont{{Wiltshire}}},
  \bibinfo{journal}{ArXiv e-prints}  (\bibinfo{year}{2011}),
  \eprint{1106.1693}.

\bibitem[{\citenamefont{{Chisari} and {Zaldarriaga}}(2011)}]{Chisari:2011iq}
\bibinfo{author}{\bibfnamefont{N.~E.} \bibnamefont{{Chisari}}}
  \bibnamefont{and}
  \bibinfo{author}{\bibfnamefont{M.}~\bibnamefont{{Zaldarriaga}}},
  \bibinfo{journal}{ArXiv e-prints}  (\bibinfo{year}{2011}),
  \eprint{1101.3555}.

\bibitem[{\citenamefont{Wiltshire}(2011)}]{Wiltshire:2011vy}
\bibinfo{author}{\bibfnamefont{D.~L.} \bibnamefont{Wiltshire}}
  (\bibinfo{year}{2011}), \eprint{1106.1693}.

\bibitem[{\citenamefont{Poisson}(2004)}]{Poisson:2003nc}
\bibinfo{author}{\bibfnamefont{E.}~\bibnamefont{Poisson}},
  \bibinfo{journal}{Living Rev.Rel.} \textbf{\bibinfo{volume}{7}},
  \bibinfo{pages}{6} (\bibinfo{year}{2004}), \eprint{gr-qc/0306052}.

\bibitem[{\citenamefont{Rasanen}(2004{\natexlab{b}})}]{Rasanen:2004js}
\bibinfo{author}{\bibfnamefont{S.}~\bibnamefont{Rasanen}},
  \bibinfo{journal}{JCAP} \textbf{\bibinfo{volume}{0411}}, \bibinfo{pages}{010}
  (\bibinfo{year}{2004}{\natexlab{b}}), \eprint{gr-qc/0408097}.

\bibitem[{\citenamefont{Paranjape and Singh}(2006)}]{Paranjape:2006cd}
\bibinfo{author}{\bibfnamefont{A.}~\bibnamefont{Paranjape}} \bibnamefont{and}
  \bibinfo{author}{\bibfnamefont{T.~P.} \bibnamefont{Singh}},
  \bibinfo{journal}{Class. Quant. Grav.} \textbf{\bibinfo{volume}{23}},
  \bibinfo{pages}{6955} (\bibinfo{year}{2006}), \eprint{astro-ph/0605195}.

\end{thebibliography}

\end{document}